\documentclass[preprint]{aastex}
\usepackage{color}
\usepackage{subfigure}
\usepackage[normalem]{ulem}
\usepackage{epsfig}
\begin{document}

\title{Measuring Organic Molecular Emission in Disks with Low Resolution \textit{Spitzer} Spectroscopy}

\author{Johanna K. Teske}
\affil{Steward Observatory, University of Arizona, 933 N. Cherry Avenue, Tucson, AZ 85721, USA}
\email{jteske@as.arizona.edu}
\author{Joan R. Najita}
\affil{National Optical Astronomy Observatory, 950 N. Cherry Avenue, Tucson, AZ 85716, USA}
\email{najita@noao.edu}
\author{John S. Carr}
\affil{Naval Research Laboratory, Code 7211, Washington, DC 20375, USA} 
\email{carr@nrl.navy.mil}
\author{Ilaria Pascucci}
\affil{Space Telescope Science Institute, 3700 San Martin Drive, Baltimore, MD 21218, USA}
\email{pascucci@stsci.edu}
\author{Daniel Apai}
\affil{Space Telescope Science Institute, 3700 San Martin Drive, Baltimore, MD 21218, USA}
\email{apai@stsci.edu}
\and
\author{Thomas Henning}
\affil{Max-Planck-Institut f\"{u}r Astronomie, K\"{o}nigstuhl\ 17, 69117\, Heidelberg, Germany}
\email{henning@mpia.de}

\clearpage

\begin{abstract}

We explore the extent to which \textit{Spitzer} IRS spectra taken
at low spectral resolution can be used in quantitative studies of 
organic molecular emission from disks surrounding low mass young stars. We use
\textit{Spitzer} IRS spectra taken in both the high and low resolution
modules for the same sources to investigate whether it is possible
to define line indices that can measure trends in
the strength of the molecular features in low resolution data. We find
that trends in HCN emission strength seen in the high resolution
data can be recovered in low resolution data.  In examining the factors
that influence the HCN emission strength, we find that the
low resolution HCN flux is modestly correlated with stellar accretion
rate and X-ray luminosity. 
Correlations of this kind are perhaps expected based on recent
observational and theoretical studies of inner disk atmospheres.
Our results demonstrate the potential of using 
the large number of low resolution disk spectra 
that reside in the \textit{Spitzer} archive to study the factors that 
influence the strength of molecular emission from disks.
Such studies would complement results for the much smaller number
of circumstellar disks that have been observed at high resolution
with IRS. 
\end{abstract}

\keywords{infrared: stars --- (stars:) circumstellar matter --- stars: pre-main sequence --- 
stars: formation --- planetary systems: protoplanetary disks}

\section{Introduction}

Circumstellar disks composed of gas and dust are ubiquitous around
forming stars and are the birthplace of planets.  Since habitable
planets are expected to form in warm inner disks ($<$ 3--4\,AU for
sun-like stars), studying this region is especially relevant to
understanding the origin and evolution of habitable planetary systems
and their diverse properties.  Interest in the origin of stars and
planets has lead to numerous studies of the gaseous components of
disks at large ($>$\,20 AU) radial distances (e.g., Dutrey et al.\
1996, 1998, 2007; Kastner et al.\ 1997; Guilloteau \& Dutrey 1998;
Thi et al.\ 2004; Semenov et al.\ 2005; Qi et al.\ 2008) as well
as warmer, solid components within $\sim$\,10\,AU of the star (e.g.,
Natta et al. 2007; Henning \& Meeus 2009; Apai \& Lauretta 2010).

Observations of the warm gas within the inner disk are also necessary
to fully understand the structure and dynamics affecting disk
evolution and planet formation (see Carr 2005; Najita et al.\ 2007a;
Millan-Gabet et al.\ 2007; Carmona 2010 for recent reviews).  When
such gas is viewed in emission from disks around T Tauri stars
(TTS), which are optically thick in the continuum at small disk
radii ($<$\,10\,AU), the emission likely originates in a temperature
inversion region at the disk surface. The very inner regions of the
gaseous disk ($< 0.3$\,AU) have been studied previously using
molecular transitions such as CO overtone emission (e.g., Carr et
al.\ 1993; Chandler et al.\ 1993; Najita et al.\ 1996, 2000, 2009;
Blum et al.\ 2004; Thi et al.\ 2005; Thi \& Bik 2005; Berthoud et
al.\ 2007) and ro-vibrational H$_{2}$O emission (e.g., Carr et al.\
2004; Najita et al.\ 2000, 2009; Thi \& Bik 2005). Observations of
CO fundamental emission (e.g., Carr et al.\ 2001; Najita et al.\
2003, 2008; Blake \& Boogert 2004; Brittain et al.\ 2007; Salyk et
al.\ 2007, 2009; Pontoppidan et al.\ 2008) and UV transitions of
H$_2$ (e.g., Valenti et al.\ 2000; Ardila \& Basri 2000; Herczeg et
al.\ 2002, 2006; Bergin et al.\ 2004) have been used to probe larger
disk radii (out to $\sim$\,1$-$2\,AU).

\begin{figure}
\includegraphics[scale=0.75]{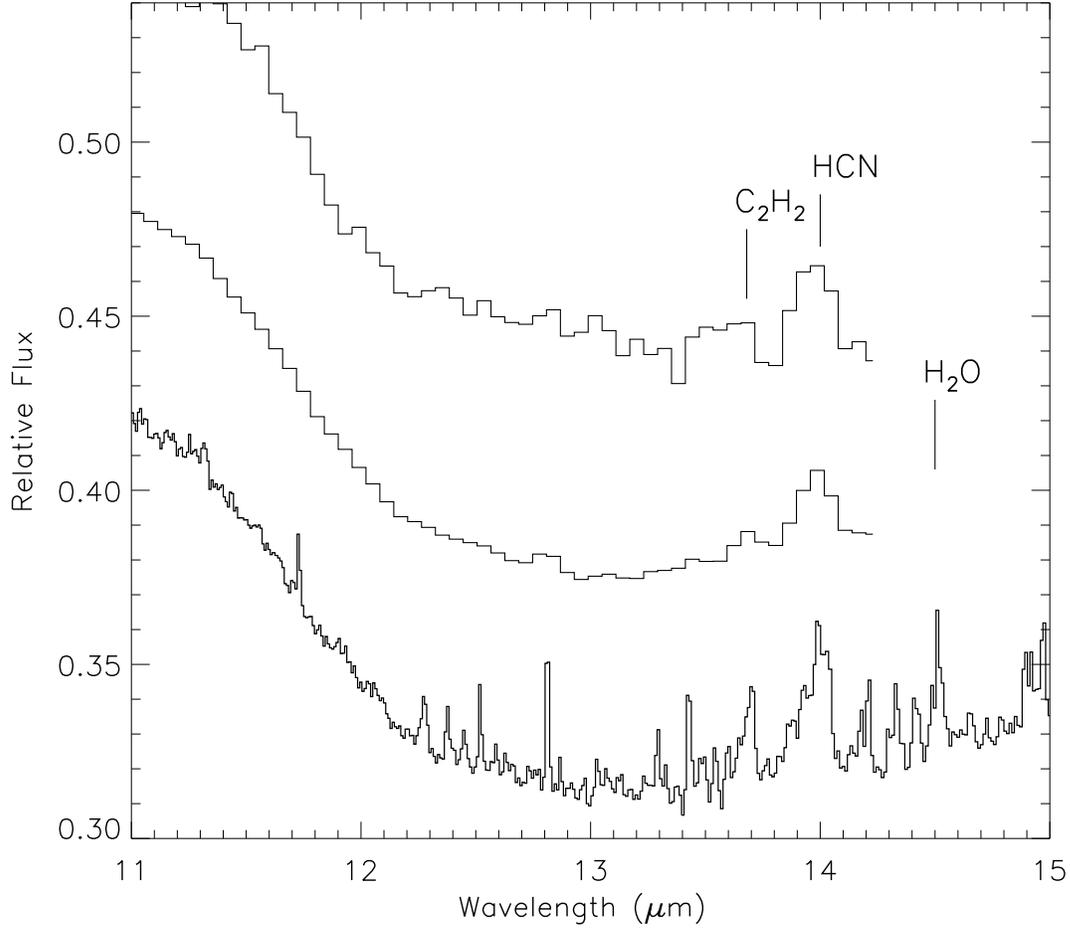}
\caption{The 11$-$15$\,\mu$m spectrum of AA Tau as observed 
in the SH ($R \sim$\,700, bottom) and SL ($R
\sim$\,100, top) modes. The middle spectrum is the SH 
spectrum smoothed to the resolution of the SL data 
and rebinned to the pixel sampling of the SL data. Several
prominent molecular features are marked with vertical lines. The
high resolution data reveal a rich emission spectrum that is common
among TTS. We show that trends in HCN emission strength in high
resolution spectra can be recovered from lower resolution data.
\label{fig1}} \end{figure}

More recently, observations of T Tauri
disks made with the high resolution ($R \sim$\,700) modules of the
Infrared Spectrograph (IRS) on board the \textit{Spitzer Space
Telescope} (Houck et al.\,2004) have revealed that simple organic molecules (HCN,
C$_{2}$H$_{2}$, CO$_{2}$) and water (Lahuis et al.\ 2006; 
Carr \& Najita 2008; Salyk et al.\ 2008) are present
in the warm inner disk region ($\lesssim$\,5\,AU). IRS observations indicate that mid-infrared molecular emission is common among TTS (Pontoppidan et
al.\ 2010; Carr \& Najita 2011; see also Pascucci et al.\ 2009 in the context of low resolution IRS data) and can be used to probe the thermal and chemical
structure of the inner gaseous disk (see Figure \ref{fig1}). 

With the cryogen of the \textit{Spitzer Space Telescope} depleted,
it is no longer possible to obtain more sensitive, mid-infrared
spectroscopy of protoplanetary disks, making the \textit{Spitzer}
archive the primary source of new information on warm disk chemistry.
However, with most of the data in the archive taken in low-resolution
mode, the question emerges: How much information regarding molecular
emission features can be extracted from the low-resolution observations?
Pascucci et al.\ (2009) previously explored this question, showing
that molecular emission could be detected in low resolution IRS
spectra of dozens TTS and lower-mass stars and brown dwarfs. They
found that HCN emission at 14\,$\mu$m was almost always brighter
than C$_{2}$H$_{2}$ emission at 13.7\,$\mu$m among T Tauri stars,
while only C$_{2}$H$_{2}$ and no HCN was detected for lower mass
stars and brown dwarfs. This led them to propose that there are
differences in the relative abundance of molecular species as a
function of stellar mass.

Here we build upon the work of Pascucci et al.\ (2009) by investigating
the extent to which we can extract quantitative information
from low resolution
\textit{Spitzer} IRS spectra of inner T Tauri disks. To do this,
we compare the molecular emission strength in a sample of high
resolution IRS spectra of T Tauri stars with similar measurements
of the same sources made in the low resolution mode of IRS. If
quantitative trends in the strength of molecular emission features
can be recovered from low resolution spectra, 
the archival \textit{Spitzer} IRS data could be used to study
the strength of molecular features in disks. 
Since, as we discuss below,
a variety of physical and chemical processes can potentially affect
the molecular emission strength, spectra of large samples of sources,
such as those available in the {\it Spitzer} archive, are a valuable
asset for demographic studies that seek to identify  the dominant
processes influencing the molecular emission from disks.
In \S 2 we describe the
data sets used in this paper. In \S 3 we present our method of
analysis and our results. The latter are explored further in \S 4.

\section{Data Sets}

For our comparison of high and low resolution data,
we examined a small set of \textit{Spitzer} IRS spectra of T Tauri
stars in the Taurus-Auriga star-forming region. The higher resolution
data were taken with IRS in the short-high mode (SH, 10$-$19$\,\mu$m, $R \sim$\,700), and the lower resolution data were taken in the
short-low mode (SL, 5.2$-$14\,$\mu$m, $R \sim$\,100). Our SH data set
was selected from classical T Tauri stars that were observed
as part of the Cycle 2 GO Program 20363 (Carr \& Najita 2008, 2011).
From 11 sources in that program, we selected a sample of ``normal''
T Tauri stars, i.e, sources with stellar accretion rates  $\lesssim
10^{-7}~ \rm M_{\odot}\rm yr^{-1}$, to avoid the influence of highly
energetic accretion processes (e.g., jets) on the spectrum. We also
excluded close binary stars (the closest companion separation in
our sample is 0.88'') since tidal interactions between the disk and
binary can disrupt or truncate the inner disk region ($< 5$\,AU).
The resulting 8 sources display mid-infrared colors that are typical
of ``normal'' TTS.  That is, they have colors that are unlike those
of transition objects. Specifically, as described in Furlan et al.\
(2006), our sample has $n_{6-13}$ between $-$1.0 and $-$0.01, and
$n_{13-25}$ between $-$0.40 and 0.17, where $n_{6-13}$ and $n_{13-25}$
are the 6$-$13\,$\mu$m and 13$-$25\,$\mu$m colors, respectively.

To compare with the 8 SH spectra, we examined SL spectra of the same objects, originally observed as part of the \textit{Spitzer} GO Program 2 (P.I. Houck). We used the reduced SL spectra from Pascucci et al.\ (2009). The observations were originally published as part of a larger data set by Furlan et al.\ (2006) using an alternative reduction procedure that they detail. Since the molecular emission features were not the focus of the latter study, those spectra were not as reliable in the 13$-$15$\,\mu$m region.

In order to determine the processes that might influence the strength of any
observed molecular emission, we also examined SL spectra of an
additional 10 sources from the Pascucci et al.\ (2009) sample that have stellar
properties similar to those of the SH sample: accretion rates within
an order of magnitude of the typical T Tauri rate $10^{-8}~\rm
M_{\odot}\rm yr^{-1}$ (Hartmann et al.\ 1998), an absence of close
companions, and normal mid-infrared colors. While the full sample
of 18 SL sources is relatively uniform in infrared spectral shape,
binarity, and spectral type, it exhibits more variety in stellar
accretion rate and X-ray luminosity (see Table \ref{tbl1}). The stellar accretion rates in Table 1 are from Hartmann et al.\ (1998) and Najita et al.\ (2007b). Najita
et al.\ adopted stellar accretion rates from several literature
sources and placed them on the same scale as Hartmann et al.,
 providing a set of comparable, consistent values. The X-ray luminosities are from the recent reanalysis of G{\"u}del et al.\ (2010) of \textit{XMM-Newton} and \textit{Chandra} observations of a large number of T Tauri stars. The X-ray luminosities are for the 0.3$-$10 keV range and have been corrected for line-of-sight absorption (G{\"u}del et al.\ 2010). We also assume a distance of 140\,pc. The properties of our full sample are described in Table ~\ref{tbl1}.
 
\begin{deluxetable}{lccccc}
\tablecaption{Our T Tauri Sample\label{tbl1}}
\small
\tablecolumns{6}
\tablenum{1}
\tablewidth{0pt}

\tablehead{\colhead{Object} & \colhead{Spectral Type$^{a}$ }  &\colhead{log(\.{M$_{*}$}/M$_{\odot}\rm yr^{-1}$)$^{c}$} & \colhead{log(L$_{\rm{X}}$/erg s$^{-1}$)$^{e}$} & \colhead{IRS Mode}} 

\startdata
AA Tau & K7 & $-$8.48 & 30.01 & SH, SL\\ 
BP Tau & K7 &  $-$7.54 &  30.16 & SH, SL\\ 
CW Tau & K3 &  $-$7.61 & $\cdots$ & SL\\ 
CX Tau & M2.5 & $-$8.97 & $\cdots$ & SL\\ 
CY Tau & M1 &  $-$8.12 & $\cdots$  & SL\\ 
DK Tau & K7 &$-$7.42 & 29.93  & SH, SL \\
DN Tau & M0 & $-$8.46 & 30.03 & SL\\ 
DO Tau & M0 & $-$6.85 & 29.37 & SH, SL\\ 
DP Tau & M0.5 & $-$7.88 & 28.99 & SL\\
DS Tau & K5 & $-$7.89 & $\cdots$ & SL\\ 
FZ Tau & M0$^{b}$  &  $-$7.32 & $\cdots$ & SL\\ 
GI Tau & K6 &  $-$8.02$^{d}$ & 29.82  & SH, SL\\ 
GK Tau & K7 & $-$8.19  &  30.09  & SH, SL\\ 
HN Tau & K5 &  $-$8.89$^{d}$  & 29.50  & SL\\
IP Tau & M0 &  $-$9.10  & $\cdots$  & SL\\ 
IQ Tau & M0.5 &  $-$7.55 & 29.50  & SL\\ 
RW Aur & K3 &  $-$7.12 &  30.21  & SH, SL\\ 
UY Aur & K7 &  $-$7.18 &  29.60  & SH, SL\\ 
   
\enddata

\tablerefs{(a) Kenyon \& Hartmann (1995), unless otherwise noted; (b) Hartigan et al.\ (1994); (c) Najita et al.\ (2007b), unless otherwise noted; (d) Hartmann et al.\ (1998); (e) G{\"u}del et al.\ (2010), corrected for line-of-sight absorption and assuming a distance of 140\,pc}

\end{deluxetable}

\clearpage 

\section{Analysis and Results}
\subsection{SH vs. SL Measurements}

As described in \S 1, Pascucci et al.\ (2009) previously showed that the $14\,\micron$ HCN feature
is almost always brighter than the $13.7\,\micron$ C$_2$H$_2$ 
feature in T Tauri spectra, making it typically the most apparent feature at low spectral resolution.
Thus, while we investigated the possibility of detecting the emission from several molecules (HCN, C$_{2}$H$_{2}$, H$_{2}$O) in the SL data, we chose to focus in this paper on HCN due to its greater detectability in our sample.

\begin{figure}
\includegraphics{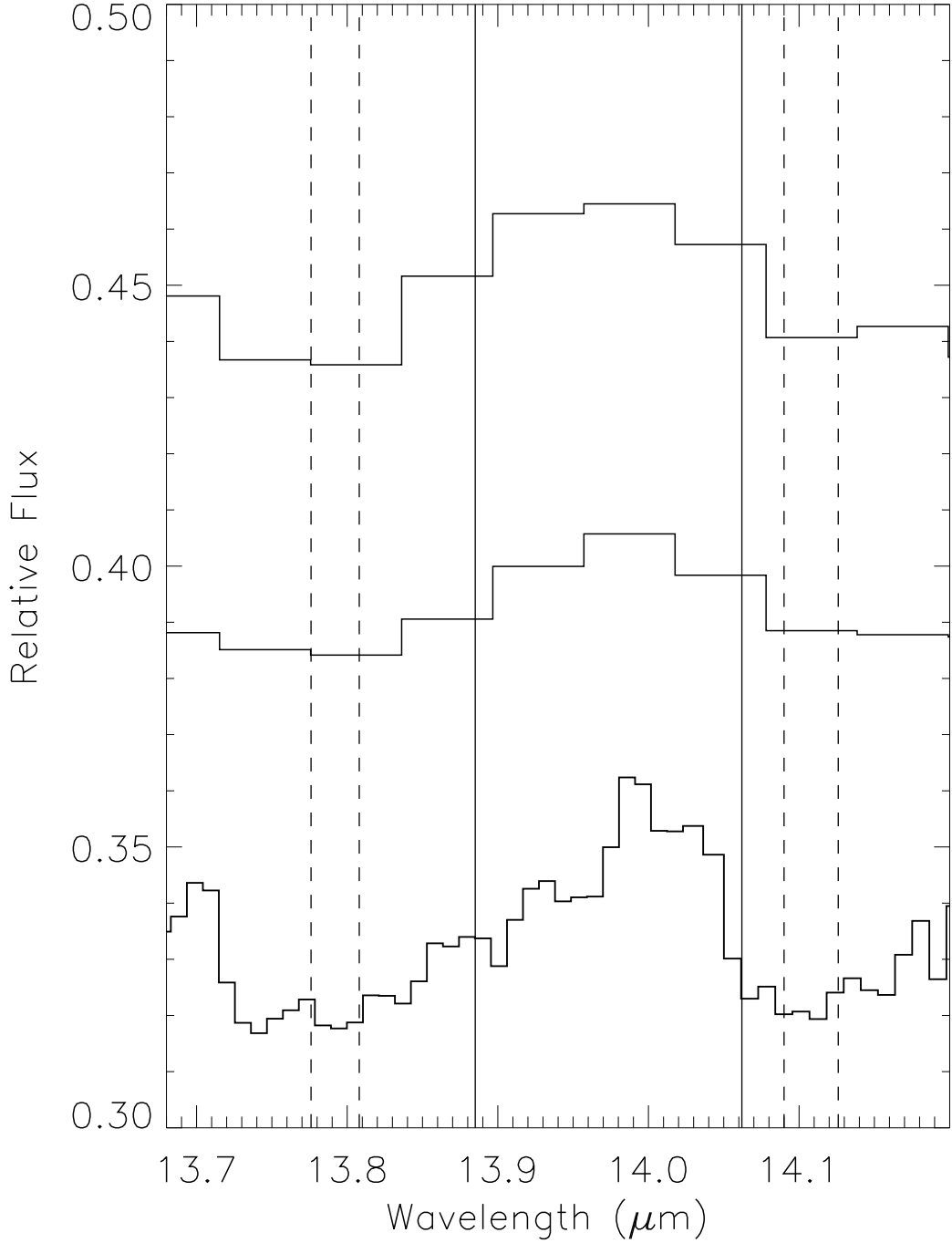}
\caption{Spectrum of AA Tau comparing SH (bottom), smoothed and resampled (middle), and SL (top) data in the region around the 14\,$\mu$m HCN feature. The dotted vertical lines indicate the left and right continuum regions, and the vertical lines define the HCN feature, as listed in \S 3.\label{fig2}}
\end{figure}

To estimate the strength of 
the HCN feature, we defined a feature index based on the structure
seen in existing SH spectra and synthetic disk emission models
(e.g., Carr \& Najita 2008) to avoid contamination from neighboring
molecular features. We selected the wavelengths
$13.885\,\mu$m and $14.062\,\mu$m to define the boundaries of the
HCN feature. To estimate the underlying continuum, we found the
average flux density in two neighboring regions, 13.776$-$13.808$\,\mu$m
and 14.090$-$14.126$\,\mu$m, assigned these values to the midpoint
of each region, and performed a linear fit to these two midpoints. We subtracted the
continuum estimate from the spectrum and summed the resulting
spectrum within the wavelength boundaries of the feature to obtain
the feature flux. The equivalent width of the feature was calculated
in a corresponding way. These values are reported in Tables 2 \&
3. In the SH spectra, the continuum
regions each span three pixels and the HCN feature spans fifteen
pixels, while in the SL spectra the continuum regions each span
less than one pixel and the HCN feature spans three pixels (see
Figure \ref{fig2}).

The errors on the SH spectra are described in Carr \& Najita\,(2011). They are derived from the average RMS pixel variation around 14$\,\micron$. To estimate the errors on the SL spectra, we performed a linear fit
to the continuum over $\sim$\,15 pixels between 13$\,\micron$ and 14.2$\,\micron$, excluding the
regions around HCN (13.885$\,\mu$m$-$ 14.062$\,\mu$m) and
C$_{2}$H$_{2}$ (13.609$\,\mu$m $-$ 13.736$\,\mu$m), and used
the standard deviation of the difference between the observed
spectrum and the fit as a measure of the pixel-to-pixel noise. We quote this measurement as our $1\sigma$ errors in Table 2.
These errors are generally smaller than those reported by 
Pascucci et al.\ (2009), who adopted an
error for each pixel based on the difference in flux observed in a
small number (2) of nod positions. 
\footnote{ This latter error estimate can be affected by flux differences in the two beam positions if the object is not equally centered in the slit
in each beam position.  Some of the spectra appeared to suffer from 
this effect as the estimated errors were often larger than the 
pixel-to-pixel differences in the final spectrum (e.g., CW Tau, CY Tau, DN Tau, GI Tau, GK Tau, IP Tau). While our errors are generally smaller than the Pascucci et al.\ (2009) errors, our adopted errors may still 
overestimate the true error.  
That is because our approach assumes that 
the true spectrum is featureless in the region used to 
estimate the pixel-to-pixel variation 
(i.e., in the regions around the HCN and C$_2$H$_2$ features), 
whereas the spectra may in fact have a rich spectrum of weaker emission 
features (Fig.~1).  
We return to this issue below.}

In Table 2 we show the SL fluxes, equivalent widths, and
errors determined using the feature and continuum regions 
defined above. 
Objects for which we have SH data are
listed in Table 3 along with their flux and equivalent width
measurements. 

To understand any difference between these two data sets, we first 
smoothed the SH
spectra to the approximate resolution of the SL spectra
($R \sim$\,100) by convolving with a Gaussian profile
and rebinned to match the SL pixel sampling. 
As these two data sets are ``contemporaneous'' (they are the same data), 
comparing them avoids any complications arising from time variability in the 
mid-infrared emission spectrum. 
We find that the fluxes and equivalent widths of the smoothed/resampled data 
are on average $\sim 50$\% of those measured for the SH data (Fig.~3a).
The lower values for the smoothed/resampled data are the result of
the neighboring line emission from water and other features 
(Fig.~1; Carr \& Najita 2008, 2011; Pontoppidan et al.\ 2010), which 
blends into a pseudo-continuum at lower spectral resolution, diluting
the HCN flux and equivalent width.  Because the neighboring line emission
can vary from source to source in both shape and strength relative
to HCN (stronger or weaker neighboring emission lines), there is 
dispersion about the $\sim 50$\% average value.  
 
We would expect that the effect of the lower spectral resolution
would lead to a similar difference between the SH measurements and
those made on the real SL data.  An additional factor in comparing
the SH data with the (non-contemporaneous) SL data is the possibility
of time variability in the HCN and/or non-HCN line emission spectrum,
which would increase the dispersion beyond that arising from the lower 
resolution alone.  This is indeed the case.  The comparison of the 
SL equivalent widths shows more dispersion than the smoothed/resampled 
data when compared against the SH data (Fig.~3b). 
Figure 3c shows that the lower average equivalent width of the 
smoothed/resampled data does indeed capture the trend of the 
reduction in the SL equivalent width.  Similar results are found 
for the HCN fluxes of the SH, smoothed/resampled, and SL data sets.

\begin{deluxetable}{lcc}
\tablecaption{HCN Short-Low Measurements\label{tbl2}}
\small

\tablecolumns{3}
\tablenum{2}
\tablewidth{0pt}
\tablehead{\colhead{Object}  &\colhead{SL HCN Flux} & \colhead{SL HCN EW}\\ 
\colhead{} & \colhead{(mJy-$\mu$m)} & \colhead{(10$^{-3}$ $\mu$m)}}
\startdata
AA Tau  & 4.00 $\pm$ 0.66 & 10.6 $\pm$ 1.77 \\
BP Tau  & 2.43 $\pm$ 0.34 & 5.77 $\pm$ 0.81 \\
CW Tau  & 3.17 $\pm$ 0.77 & 4.15 $\pm$ 1.02 \\
CX Tau  & $-$0.178 $\pm$  0.53 & $-$1.16 $\pm$  3.35 \\
CY Tau  & $-$0.170 $\pm$ 0.54 & $-$1.47 $\pm$ 4.60 \\
DK Tau  & 1.75 $\pm$ 0.70 & 1.96 $\pm$ 0.78 \\
DN Tau  & 2.20 $\pm$  0.39 & 7.04 $\pm$ 1.28 \\
DO Tau  & 1.87 $\pm$ 1.27 & 1.19 $\pm$ 0.81 \\
DP Tau  & 0.608 $\pm$ 0.66 & 1.03 $\pm$ 1.11 \\
DS Tau  & 2.33 $\pm$ 0.23 & 9.65 $\pm$ 0.99 \\
FZ Tau  & 4.37 $\pm$ 1.38 & 4.93 $\pm$ 1.58 \\
GI Tau  & 2.99 $\pm$ 0.62 & 4.09 $\pm$ 0.85 \\
GK Tau  & $-$1.13 $\pm$ 0.77 & $-$1.37 $\pm$ 0.93 \\
HN Tau  & 0.783 $\pm$ 0.59 & 0.992 $\pm$ 0.75 \\
IP Tau  & $-$0.964 $\pm$ 0.69 & $-$5.26 $\pm$ 3.75 \\
IQ Tau  & 2.02 $\pm$ 0.62 & 6.07 $\pm$ 1.90 \\
RW Aur  & 5.51 $\pm$ 1.34 & 4.32 $\pm$ 1.06 \\
UY Aur  & 2.62 $\pm$ 1.30 & 0.871 $\pm$ 0.43 \\
\enddata
\end{deluxetable}

\begin{deluxetable}{lcccc}
\tablecaption{HCN Short-High and Smoothed \& Resampled Measurements\label{tbl3}}
\tablewidth{0pt}
\small
\tablecolumns{5}
\tablenum{3}

\tablehead{\colhead{Object} & \colhead{SH HCN Flux} &  \colhead{SH HCN EW} & \colhead{SM+RS HCN Flux} &  \colhead{SM+RS HCN EW}  \\ 
\colhead{} & \colhead{(mJy-$\mu$m)} & \colhead{(10$^{-3}$ $\mu$m)} & \colhead{(mJy-$\mu$m)} & \colhead{(10$^{-3}$ $\mu$m)}}
\startdata
AA Tau  & 4.43  $\pm$ 0.08 & 13.9  $\pm$ 0.25 & 2.52  $\pm$  0.06 & 7.72 $\pm$ 0.18 \\
BP Tau  & 4.31  $\pm$ 0.08 & 11.6  $\pm$ 0.21& 2.35  $\pm$  0.06 & 6.21 $\pm$ 0.15 \\
DK Tau  & 5.01  $\pm$ 0.15 & 6.59  $\pm$ 0.19 & 2.20  $\pm$ 0.10 & 2.86 $\pm$ 0.13 \\
DO Tau  & 1.32  $\pm$ 0.25 & 0.653  $\pm$ 0.12 & $-$0.676  $\pm$ 0.12 & $-$0.333 $\pm$ 0.09\\
GI Tau  & 4.69  $\pm$ 0.12 & 6.18  $\pm$ 0.15 & 2.47  $\pm$ 0.08 & 3.22 $\pm$ 0.11\\
GK Tau  & 0.850  $\pm$ 0.15  & 1.07  $\pm$ 0.15 & 0.104  $\pm$ 0.09 & 0.130 $\pm$ 0.11 \\
RW Aur  & 9.36  $\pm$  0.25 & 6.32  $\pm$ 0.17 & 3.97  $\pm$ 0.18 & 2.64 $\pm$ 0.12 \\
UY Aur  & 5.95  $\pm$ 0.34 & 2.24  $\pm$ 0.12 & 1.56  $\pm$ 0.24 & 0.58 $\pm$  0.09 \\
\enddata
\end{deluxetable}

\newcommand{\txw}{\textwidth}
\begin{figure}[ht!]
\centering

  {\epsfig{file=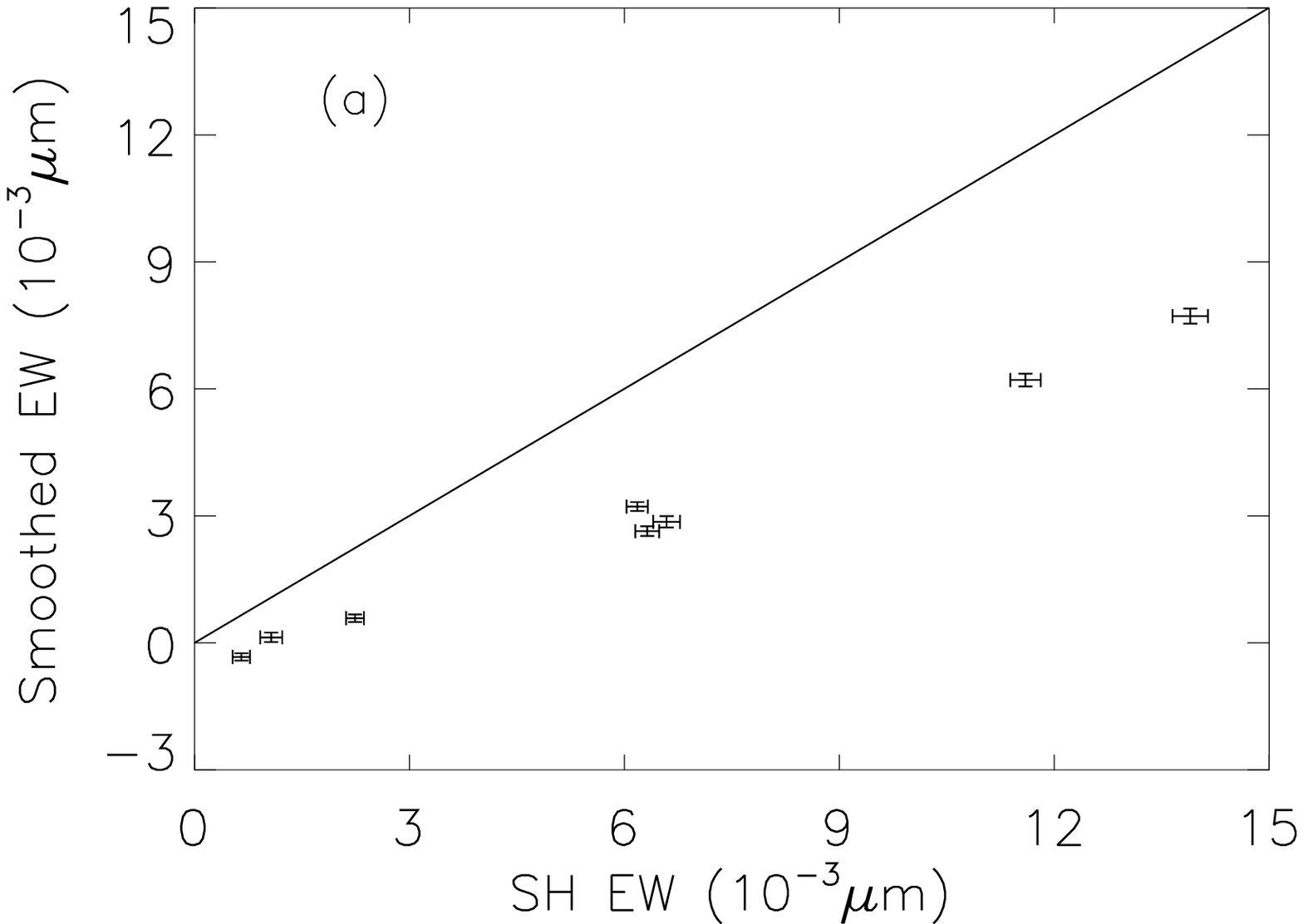,width=0.5\txw,clip=}}
\par   {\epsfig{file=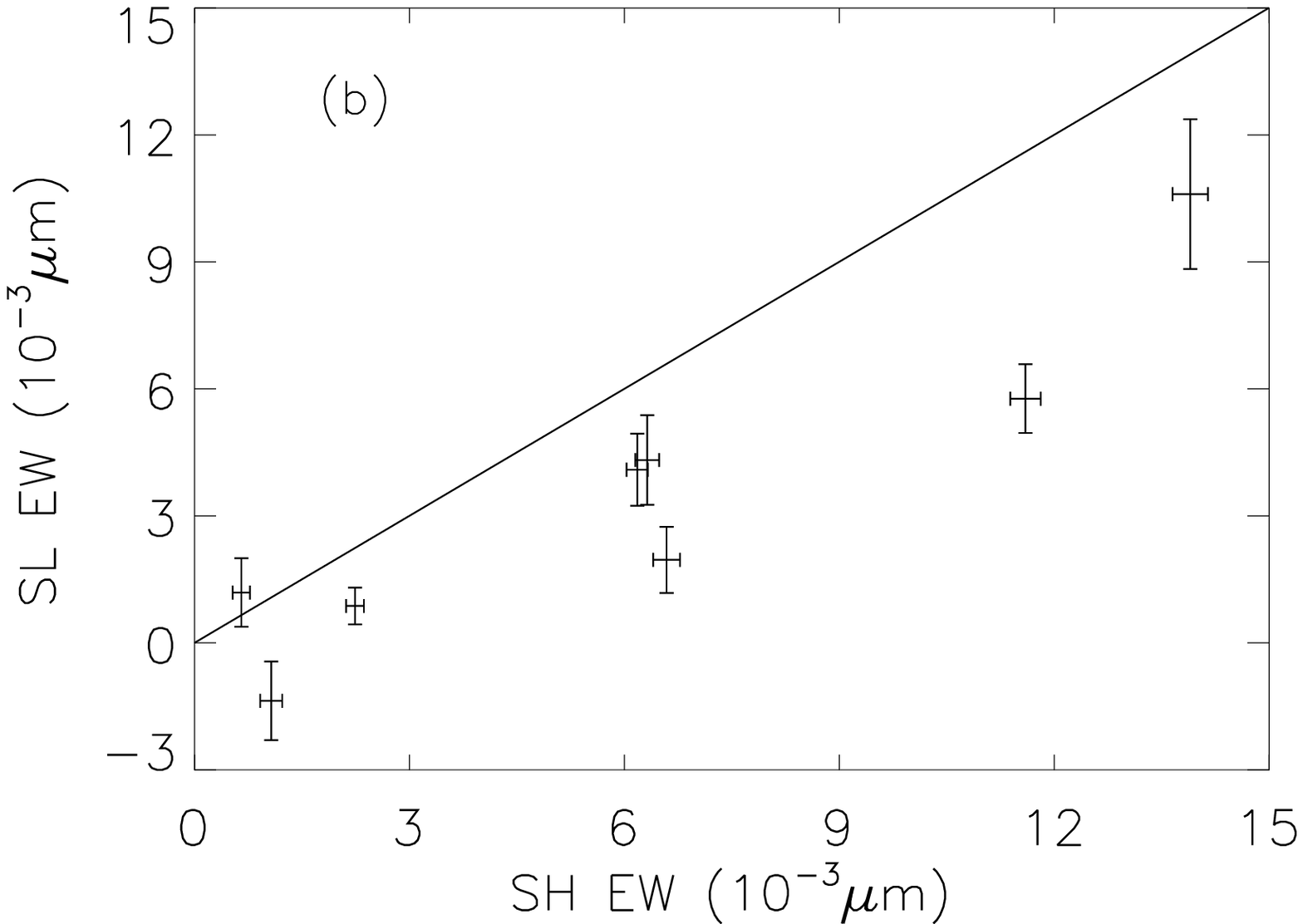,width=0.5\txw,clip=}}
\par   {\epsfig{file=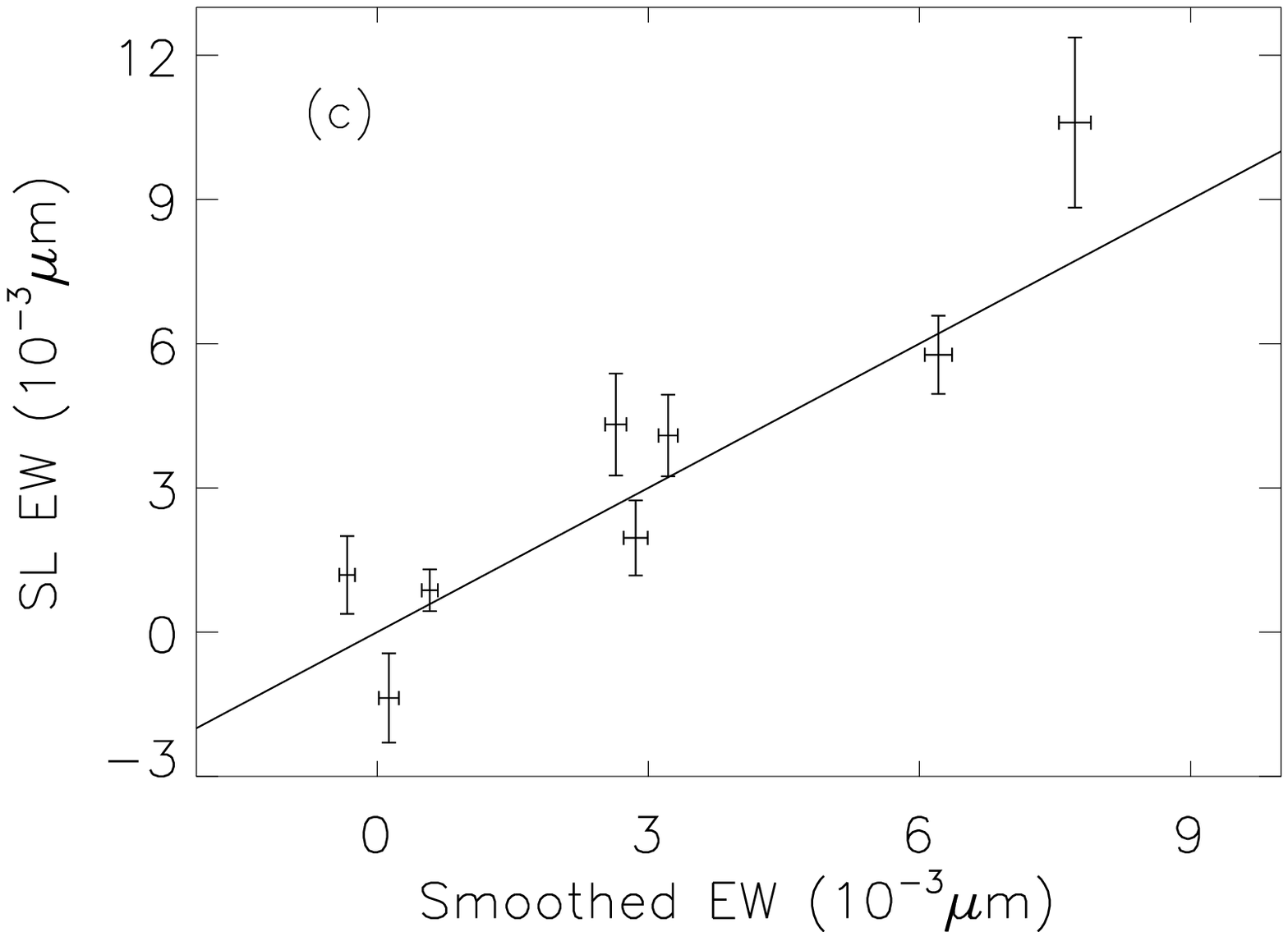,width=0.5\txw,clip=}}
\caption{The comparison SH, smoothed/resampled and SL HCN equivalent widths (Tables 2 and 3). A unity line is shown for reference in each plot.}
\label{fig3}
\end{figure}

The HCN equivalent width and flux measurements from the SH and
SL data are well correlated (Figure \ref{fig3} and Table \ref{tbl4}). 
To assess the significance of the apparent trends, we use two
correlation coefficients, Kendall's rank correlation coefficient,
$\tau_{Kendall}$, and Pearson's linear correlation coefficient,
\textit{r}. The former, $\tau_{Kendall}$, is a non-parametric
statistic that measures the degree of correlation between two
variables; values close to unity signify a tighter correlation,
while values close to 0 signify no correlation. Our calculated
$\tau_{Kendall}$-values are all $\ge$\, 0.59. The two-sided \textit{P}
values that correspond to $\tau_{Kendall}$, $P{_\tau}$, represent
the confidence levels of the coefficient -- a smaller \textit{P} value
indicates a lower probability of a false conclusion. Pearson's
\textit{r}-value  measures how closely two parameters fit a linear
relationship (assuming the parameter distributions are normal). The
closer $\left|r\right|$ is to unity, the more linear the relationship.
Our calculated \textit{r}-values are all $\ge$\, 0.80, signifying
a near-linear correlation. We also calculate $p_{rand}$ (as a \%),
the probability that our measurements are randomly distributed (and
thus uncorrelated). The calculated $p_{rand}$ values, $\sim$\,1.8\%
and 7.0\% for the equivalent width and flux relations, respectively,
indicate that it is highly unlikely that our measurements are
randomly distributed. These statistics for the above trends are
shown in Table \ref{tbl4}. As we suspect and the figures indicate,
the trends we find are statistically significant.

\begin{deluxetable}{lccccc}
\tablecaption{Correlation Between SH and SL HCN Emission \label{tbl4}}
\tablewidth{0pt}
\tablenum{4}
\tabletypesize{\small}
\tablecolumns{6}
\tablehead{\colhead{Parameters} & \colhead{n\tablenotemark{a}} & \colhead{\textit{r}\tablenotemark{b}} & \colhead{$\tau_{Kendall}$\tablenotemark{c}} & \colhead{P$_{\tau}$\tablenotemark{d}} & \colhead{$p_{rand}$ (\%)\tablenotemark{e}}} 
\startdata

SH vs. SL EW & 8 & 0.904 & 0.714 & 0.019  & 1.77 \\
SH vs. SL Flux & 8 & 0.797 & 0.590 & 0.108 & 7.01 \\
Smoothed vs. SL EW & 8 & 0.908 & 0.714 & 0.019  & 1.64 \\
\enddata

\tablenotetext{a}{\textit{The number of objects used for calculation of the statistic.}}
\tablenotetext{b}{\textit{Pearson's r linear correlation coefficient, a measure of how closely two variables fit a linear relationship. $\left|r\right|$ values closer to 1 indicate better correlation.}}
\tablenotetext{c}{\textit{Kendall's $\tau$ rank statistic, a measure of the degree of correlation between two parameters that does not assume normally distributed data. The closer $\left|\tau\right|$ is to 1, the better the correlation.}}
\tablenotetext{d}{\textit{Two-sided \textit{P} value, the probability (assuming no correlation) of obtaining a result at least as extreme as the result that is actually observed. The lower the \textit{P} value, the higher the probability of correlation..}} 
\tablenotetext{e}{\textit{Probability of getting r from a random distribution of size n.}}

\end{deluxetable}

\begin{deluxetable}{lcccccccc}
\tablenum{5}
\tabletypesize{\scriptsize}
\tablecaption{Correlations Between Stellar Parameters \& SL HCN Emission \label{tbl5}}
\tablewidth{0pt}

\tablecolumns{9}
\tablehead{\colhead{Parameters}                                     &
           \colhead{Points Rejected}                      &
	   \colhead{n\tablenotemark{a}}                             &
	   \colhead{\textit{r}\tablenotemark{b}}                    & 
	   \colhead{$\tau_{Kendall}$\tablenotemark{c}}                &
           \colhead{P$_{\tau}$\tablenotemark{d}}                      & 
           \colhead{$p_{rand}$ (\%)\tablenotemark{e}}              &
           \colhead{$\chi^{2}$ } &
           \colhead{\textit{q}}}
\startdata
log(\.{M$_{*}$}/M$_{\odot}\rm yr^{-1}$) vs. SL Flux -- initial fit  & none & 18 & 0.534 & 0.386 & 0.028 & 8.42 & 1.17  & 0.280  \\
log(\.{M$_{*}$}/M$_{\odot}\rm yr^{-1}$) vs. SL Flux -- final fit & 2 & 16 & 0.655 & 0.567 & 0.178  & 2.87 & 0.753  &  0.721\\
\hline
log(L$_{\rm{X}}$/erg s$^{-1}$) vs. SL Flux -- initial fit  & none & 12 & 0.403 & 0.382 & 0.099  &  34.19 & 1.37   & 0.186 \\
log(L$_{\rm{X}}$/erg s$^{-1}$) vs. SL Flux -- final fit  & 1 & 11 & 0.648 & 0.587 & 0.016  & 9.80 & 0.676  & 0.731 \\
\hline
Spectral Type vs. SL Flux -- initial fit  & none & 18 & $-$0.541 & $-$0.405 & 0.028  & 7.26 &  3.34 & 0.00 \\
Spectral Type vs. SL Flux -- final fit  & 1 & 17 & $-$0.564 & $-$0.485 & 0.010  & 6.19 &  3.12  &  0.00 \\
\hline
\hline
\hspace{5pt}
Spectral Type vs. log(L$_{\rm{X}}$/erg s$^{-1}$)  & none & 12 & $-$0.464 & $-$0.355 & 0.150  & 26.34 & 1.33   & 0.205 \\
Spectral Type vs. log(\.{M$_{*}$}/M$_{\odot}\rm yr^{-1}$)  & none & 18 & $-$0.269 & $-$0.154 & 0.417  & 52.36 & 1.48   & 0.096 \\
log(L$_{\rm{X}}$/erg s$^{-1}$) vs. log(\.{M$_{*}$}/M$_{\odot}\rm yr^{-1}$) & none & 12 & $-$0.084 & 0.015 & 1.00 & 67.47 & 1.62   & 0.095 \\
\enddata
\tablecomments{See description of parameters in Table \ref{tbl4} and in text.}

\end{deluxetable}

To summarize, while the SL measurements do not recover the HCN flux 
of the SH spectra, our results suggest that studies
using SL spectra can recover quantitative trends in molecular emission strength
seen in higher resolution observations. 
The SL HCN measurements may therefore track the SH HCN measurements 
well enough to reveal interesting trends when compared with other T Tauri 
properties.  We explore this possibility in the next section.

Although we refer to our SL measurements as ``fluxes'' and
``equivalent widths'', it is more useful to think of these quantities
as \textit{line} indices.  The index can be positive (e.g., if there is HCN
emission) or negative.  The latter could occur either if there is
either true absorption (e.g., as in IRS 46; Lahuis et al.\ 2006) or
emission from other features in the ``continuum'' regions that are
used to define the index.

In addition to the HCN emission feature, we also attempted a similar
analysis for C$_{2}$H$_{2}$ ($\sim$\,13.7$\,\mu$m) and an H$_{2}$O
feature at $\sim$\,12.4$\,\mu$m. We were unable to recover with the 
SL data emission strength trends seen in the SH data for these features, probably 
because they are weaker than HCN in spectra of T Tauri stars 
(Pascucci et al.\ 2009).
We note that greater success may
be possible with data analysis techniques more sophisticated than
those used here. We also note that when we performed the same analysis using the Furlan et al.\ (2006) reduction of the SL data we did not find a significant
correlation between SL and SH emission strengths, demonstrating
that the specific data reduction procedure for the SL data can
influence the ability to recover trends in SH data.

\subsection{Variation in HCN Feature Strength}

In our sample of SL spectra, the HCN flux varies
from non-detections (below $\sim$\,1 mJy-$\mu$m) to over 5 mJy-$\mu$m, and the HCN
equivalent width varies over approximately an order of magnitude
(see Table 2). What causes the strength of the HCN feature to differ in these
systems? Although the sources have many similar properties (e.g.,
they have similar stellar masses and spectral types), the stellar
accretion rate (\.M$_*$) and X-ray luminosity (L$_{\rm{X}}$)
do vary across the sample, 
 as may other physical properties not described here.
To investigate whether stellar accretion rate and X-ray luminosity 
play a role in determining the HCN emission strength,
we compared the HCN fluxes of the sources in the SL sample with
their values of \.M$_*$ and L$_{\rm{X}}$ from the literature (Table 1). 

\begin{figure}[t!]
\centering
\mbox{   {\epsfig{file=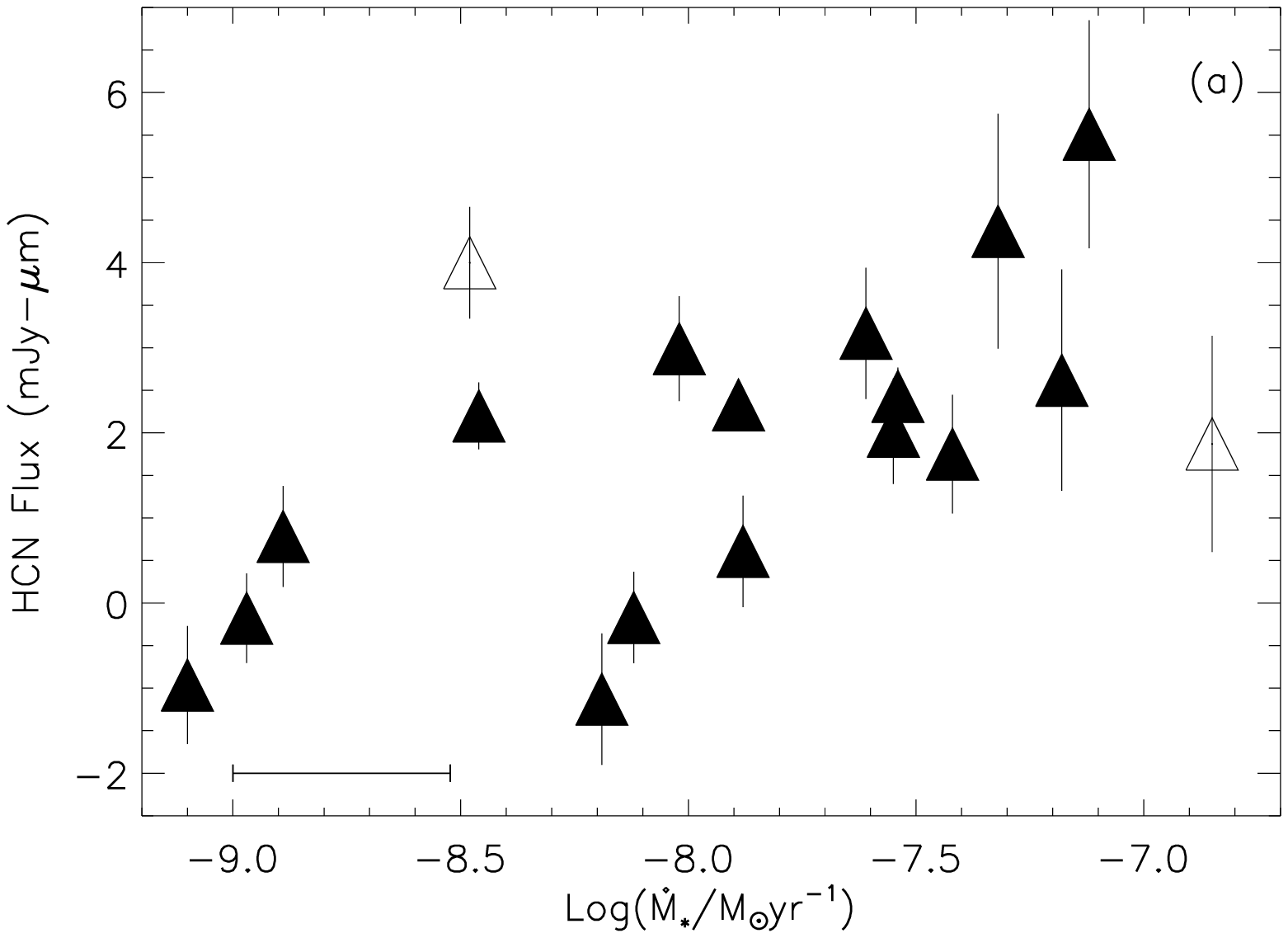,width=0.5\txw,clip=}}
   {\epsfig{file=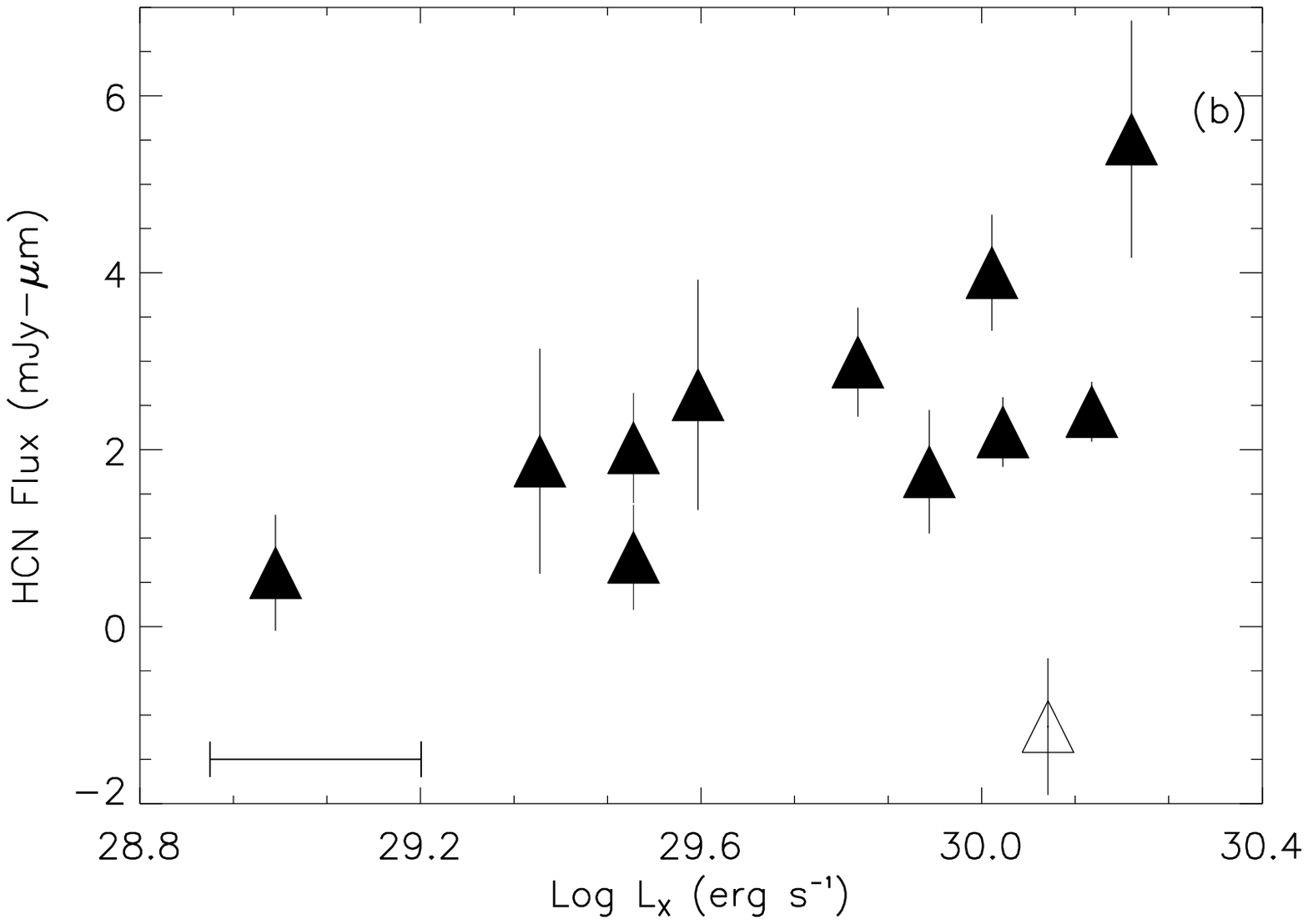,width=0.5\txw,clip=}}}
\par

\mbox{   {\epsfig{file=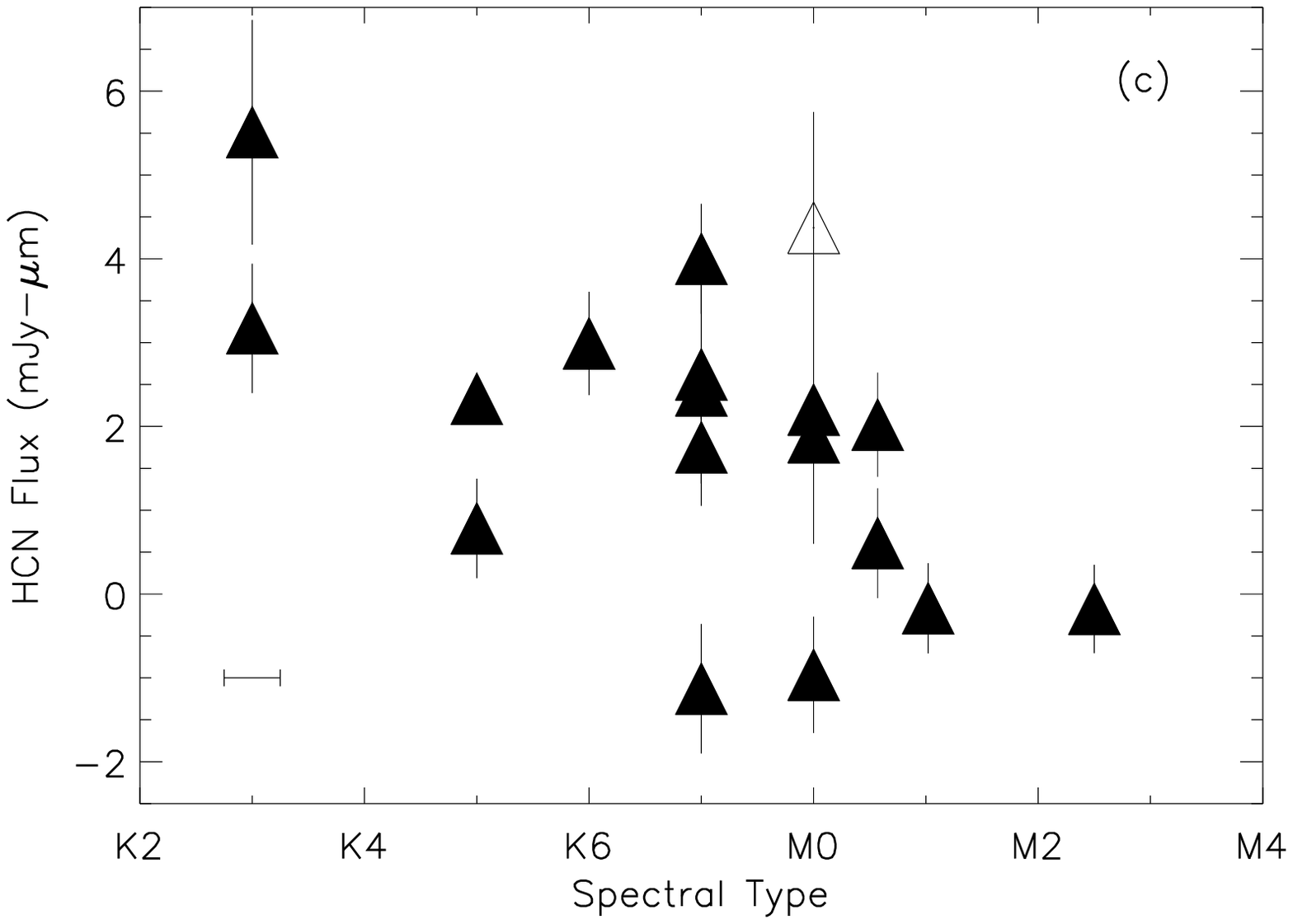,width=0.5\txw,clip=}}
   {\epsfig{file=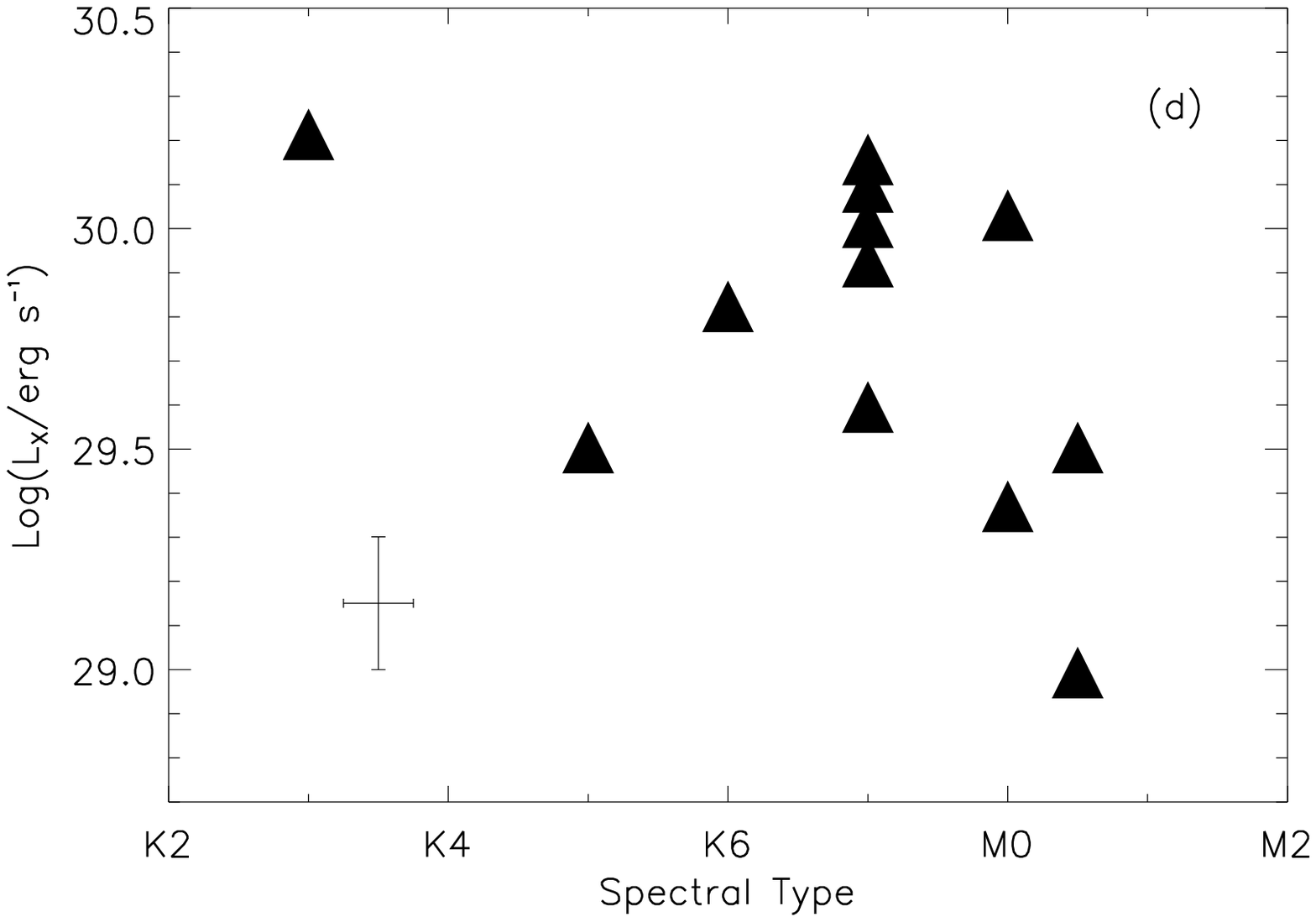,width=0.5\txw,clip=}}}
\par
\mbox{   {\epsfig{file=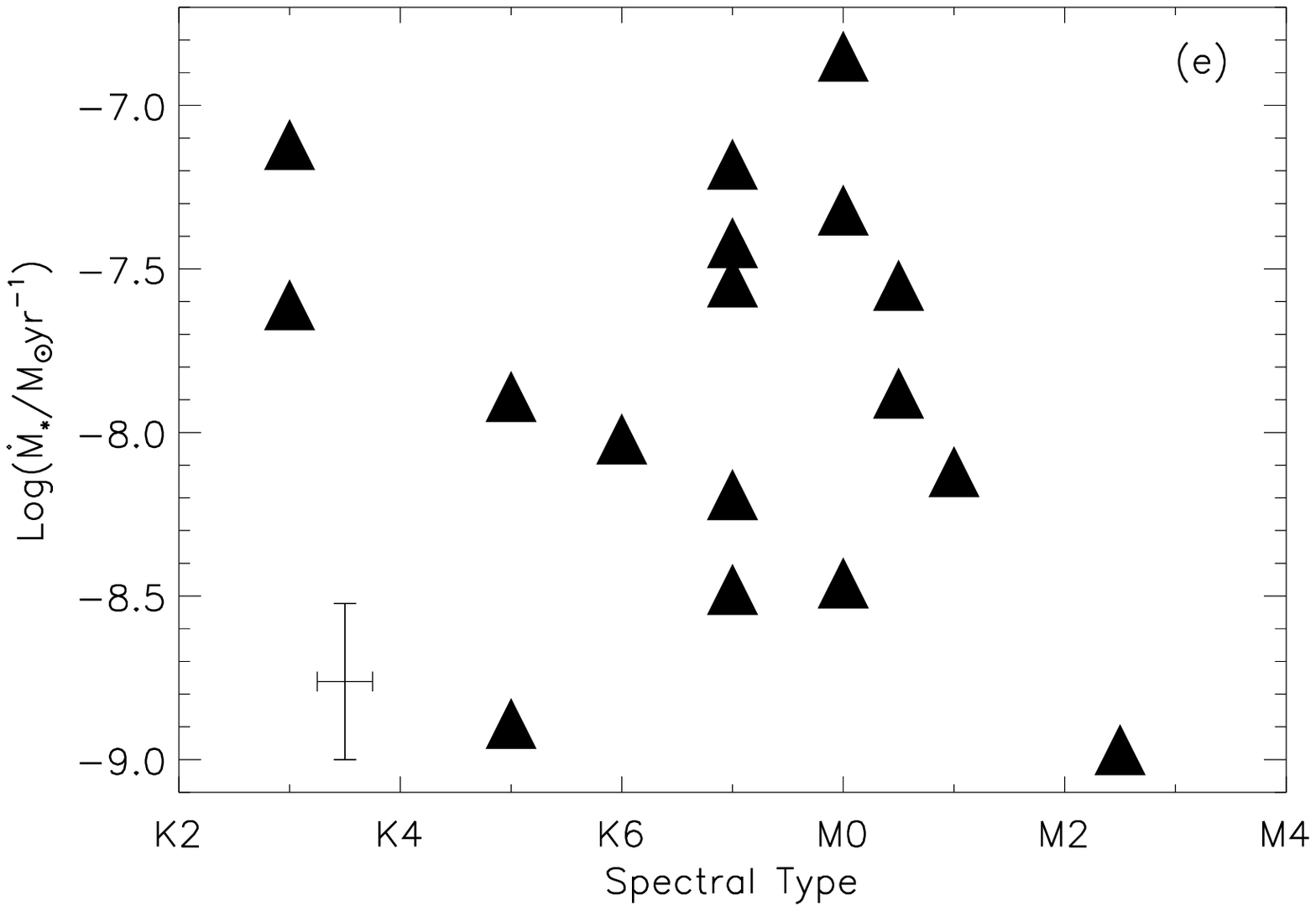,width=0.5\txw,clip=}} 
   {\epsfig{file=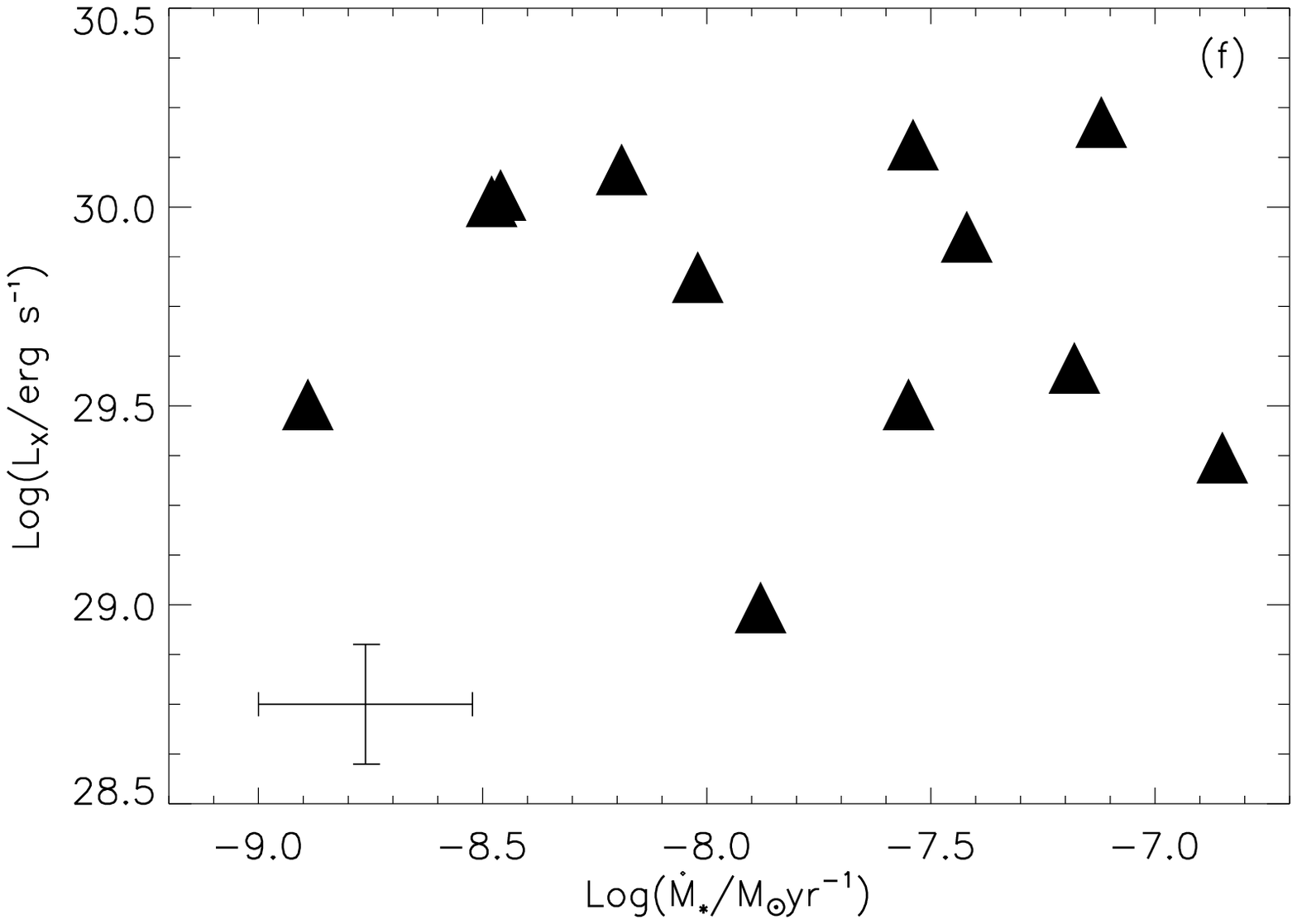,width=0.5\txw,clip=}}}
\caption{The comparison of stellar parameters and SL HCN flux. Open triangles designate outliers identified by iterative rejection. (a) SL HCN flux versus stellar accretion rate (\.{M$_{*}$}). (b) SL HCN flux versus stellar X-ray luminosity (L$_{\rm{X}}$). (c) SL HCN flux versus spectral type. (d) L$_{\rm{X}}$ versus spectral type. (e) \.{M$_{*}$} versus spectral type. (f) L$_{\rm{X}}$ versus \.{M$_{*}$}.}
\label{fig_fp}
\end{figure}

In Figure \ref{fig_fp}, panels (a), (b), and (c) plot SL HCN flux against stellar
accretion rate, stellar X-ray luminosity, and spectra type,
respectively. Panels (d), (e), and (f) plot these three quantities --
accretion rate, X-ray luminosity, and spectral type -- against each
other. The distribution of points suggests possible trends between
SL HCN flux and the quantities in Fig.\,\ref{fig_fp}a, b, c, although these
trends, if they exist, are not extremely tight. The lack of a tight correlation is perhaps not
surprising since many physical and chemical processes (e.g., heating
that is unrelated to accretion, chemical synthesis, photodestruction,
excitation conditions) can potentially affect the strength of any
given molecular emission feature. As a result, outliers in any trend
are to be expected, e.g., if some systems have managed to synthesize
more or less HCN. We therefore employed the following simple rejection scheme  
when examining our data for possible trends: 
we performed a weighted linear fit, including uncertainties in
both the x- and y-directions, to all of the data in Fig.\,\ref{fig_fp}a, b,
and c and iteratively rejected the top one to two outliers, 
all of which were above 3.3-$\sigma$. The outliers
are plotted as open triangles in Fig.\,\ref{fig_fp}, and a summary of the fit
statistics is given in Table \ref{tbl5}.
Table \ref{tbl5} also reports the reduced $\chi^2$ of the linear
fit and $q$, the probability that a correct model would give a $\chi^2$ value
equal to or larger than the observed $\chi^2$.

In the case of Fig.\,\ref{fig_fp}a, where we plot SL HCN flux versus stellar
accretion rate, the  Pearson's \textit{r}-value associated with all
of the data points shown is 0.53 and the $\tau_{Kendall}$ value is
0.39 (see Table \ref{tbl5}). Rejection of the top two
outliers at 3.6$\,\sigma$ and 3.8$\,\sigma$ (open symbols) resulted
in a Pearson's \textit{r}-value associated with the remaining
data points of 0.66 and the $\tau_{Kendall}$ value of 0.57 (see
also Table \ref{tbl5}), suggesting a potential positive correlation
between stellar accretion rate and HCN flux.

Even with outlier rejection, there is still significant scatter,
which is perhaps to be expected, as discussed above. In addition,
the difficulty in determining precise veiling and bolometric
corrections likely introduces systematic uncertainty in stellar
accretion rate measurements, as discussed by Hartigan et al.\ (1991)
and Gullbring et al.\ (1998). These authors also note that time
variability, as a result of intrinsic fluctuation in the accretion
rate or the modulation of a nonaxisymmetric magnetosphere, can
contribute to the uncertainty; they suggest a cumulative uncertainty
of $\sim$\,3 in stellar accretion rate (Hartigan et al.\ 1991;
Gullbring et al.\ 1998). We represent this uncertainty by the
horizontal bar in the lower left corner of Fig.\,\ref{fig_fp}a.

For Fig.\,\ref{fig_fp}b, which shows SL HCN flux versus stellar X-ray luminosity,
the associated Pearson's \textit{r}-value for all of the data points
is 0.40 and the $\tau_{Kendall}$ value is 0.38 (see Table \ref{tbl5}).
Rejection of the top outlier at 3.3$\,\sigma$ (open symbol)
resulted in a Pearson's \textit{r}-value associated with the
remaining data points of 0.65 and the $\tau_{Kendall}$ value of
0.59 (see also Table \ref{tbl5}), suggesting a potential positive
correlation between stellar X-ray luminosity and HCN flux. The
larger $p_{rand}$ and P${_\tau}$ for these data (compared to those
shown in Fig.\,\ref{fig_fp}a or 4c; see Table \ref{tbl5}) are partly a result
of the smaller sample size \textit{n} (12 versus 18 objects). Some
of the scatter in this plot is likely the result of variability in
L$_{\rm{X}}$. G{\"u}del et al.\ 2010 notes that the range of
uncertainty in X-ray flux determination is dominated by variability
on various time scales, and (apart from singular flares) is typically
characterized by flux variations within a factor of two from low
to high levels. We represent this uncertainty by the horizontal
error bar in the lower left corner Fig.\,\ref{fig_fp}b.

Fig.\,\ref{fig_fp}c shows SL HCN flux versus stellar spectral type.
The associated Pearson's \textit{r}-value
for all of the data points is $-$0.54 and the $\tau_{Kendall}$ value
is $-$0.41 (see Table \ref{tbl5}). Rejection of the top
outlier 3.5$\,\sigma$ (open symbol) resulted in a Pearson's
\textit{r}-value associated with the remaining data points of $-$0.56
and the $\tau_{Kendall}$ value of $-$0.49 (see also Table
\ref{tbl5}). 
An estimated spectral type error of 0.5 subclass is much smaller than the 
dispersion of the points.
While the statistics suggest a possible negative
correlation between spectral type and HCN flux, it seems unlikely
that spectral type (and therefore stellar temperature) directly
affects the HCN flux from the disk; while the HCN flux in our sample
varies over almost an order of magnitude, the range of spectral
types we studied is relatively narrow, spanning $\sim 1400$\,K in
temperature.

Fig.\,\ref{fig_fp}d may shed some light on this issue. It shows that within our
sample, X-ray luminosity decreases on average with later spectral
type. The associated Pearson's \textit{r}-value for all the objects
plotted is  $-$0.46, and the $\tau_{Kendall}$ value is
$-$0.36 (see Table \ref{tbl5}). This modest correlation in our
sample is also supported by larger samples of pre-main sequence
stars (e.g., Winston et al.\ 2010; Preibisch et al.\ 2005); our
examination of those data show a similar decrease in X-ray luminosity
with later spectral type. This trend between X-ray luminosity and
spectral type could be explained as a consequence of the rough
proportionality between L$_{\rm{X}}$ and L$_{*}$ in pre-main sequence
stars, with L$_{\rm{X}}$/L$_{*}\sim$\,10$^{-4}-$10$^{-3}$ (Telleschi
et al.\ 2007; Preibisch et al.\ 2005). Among stars in Myr-old
populations such as those in our sample, L$_{*}$ also decreases
with later spectral type (Stelzer \& Neuh{\"a}user 2001; Preibisch
et al.\ 2005; Winston et al.\ 2010), so L$_{\rm{X}}$ would also be
expected to decrease with later spectral type in our sample. Thus,
the trend in Fig.\,\ref{fig_fp}c may not reflect a fundamental relationship
between HCN flux and spectral type, but instead results from the two
underlying relations between L$_{\rm{X}}$ and HCN flux (Fig.\,\ref{fig_fp}b)
and L$_{\rm{X}}$ decreasing with later spectral type (Fig.\,\ref{fig_fp}d).

Another possibility is that the luminosity associated with accretion (L$_{\rm{acc}}$) is decreasing with later spectral type and this is what drives the trend of HCN flux with spectral type. The average accretion rate is known to decrease with decreasing mass (later spectral type), but the spread at any given mass is $\sim$\,two orders of magnitude (Muzerolle et al.\ 2005). In Fig.\,\ref{fig_fp}e, we plot stellar accretion rate versus spectral type. There is no strong correlation (see Table \ref{tbl5}) within the narrow range of spectral type of our sample, consistent with Muzerolle et al.\ (2005). In Fig.\,\ref{fig_fp}f, we plot the stellar X-ray luminosity versus the stellar accretion rate. The comparison also shows no correlation (see Table \ref{tbl5}).

Because our data set is small (and our analysis methods explorative), larger samples of IRS spectra are needed to confirm that any trends exist and test whether any of the fits proposed are reasonable representations of the trend. Our
sample is artificially sparse at high accretion rates due to the
difficulty in measuring HCN emission from low resolution spectra
of high-accretion sources; their enhanced continuum flux reduces
the contrast of emission features above the continuum. Thus
it would be useful to expand the sample to include more sources
covering the same range of stellar accretion rates as well as a
larger range of accretion rates. If HCN flux and stellar accretion rate are correlated, we would expect that sources with accretion rates $<10^{-9} ~\rm M_{\odot}\rm yr^{-1}$ would have low to undetectable HCN fluxes. Similarly, we would expect that sources with X-ray luminosities below $\sim 6.3\times 10^{28}$ erg s$^{-1}$
would not show detectable HCN, and that sources with X-ray luminosities above 
$\sim 2.5 \times 10^{30}$ erg s$^{-1}$ might continue to show enhanced HCN emission with increasing X-ray flux. 

\vfill
\clearpage

\section{Discussion}

We find that SH and SL measurements of the $14~\mu$m HCN feature are 
correlated in our small sample of T Tauri stars. 
Our results
support the work of Pascucci et al.\ (2009), who used these SL spectra
as part of their larger sample to deduce the differences between
gaseous disks surrounding T Tauri stars and those surrounding lower 
mass stars and brown dwarfs. That study showed a prominent difference in the relative detection rates of HCN and C$_{2}$H$_{2}$ between the two samples, with HCN detected more commonly in TTS than in the lower mass objects. The median spectra they created of samples of T Tauri stars and the lower mass objects showed that the flux ratio
of HCN to C$_2$H$_2$ is $\sim$\,3 for T Tauri stars and much lower, 
$\sim$\,0.2, for the lower mass objects. Our results show that such comparisons can be extended to comparisons of HCN feature fluxes in the spectra of individual objects.

We also find potential trends between the SL HCN flux
index and stellar accretion rate, X-ray luminosity, and stellar spectral type. With respect to the potential trend with stellar accretion rate, a 
similar relation between CO fundamental emission and stellar
accretion rate has been reported in TTS and Herbig Ae-Be stars 
(Najita et al.\ 2003; Brittain et al.\ 2007). These authors suggest
that a correlation between CO emission and accretion rate would be
expected if accretion-related processes heat the disk atmosphere. In
a related study of transition objects, Salyk et al.\ (2009) report that the
sources in their sample that show inner-disk CO fundamental emission
have higher accretion rates. The sources that display CO fundamental
emission also display Pf$\beta$ emission, which is
moderately correlated with the accretion diagnostic H$\alpha$. Accretion-related processes could strengthen the HCN emission by enhancing the temperature, and/or the HCN abundance, in the disk atmosphere.

The effect of accretion-related heating on disk molecular emission has been studied by Glassgold et al.\ (2004, 2009). They proposed two sources of mechanical heating in the disk atmosphere: viscous accretion, possibly generated by the magnetorotational instability (MRI; Stone et al.\ 2000), and stellar wind interaction with the disk surface (Glassgold et al.\ 2004). Glassgold et al.\ (2009) invoked mechanical heating, due to one or both of these sources, in addition to the formation of H$_2$ on warm grains, to explain the large column densities of warm H$_{2}$O that are observed in emission in disk atmospheres. Glassgold et al.\ (2009) determined that these processes can increase the thickness of the warm water column to the extent reported by Carr \& Najita (2008) and Salyk et al.\ (2008). If mechanical heating does affect the thermal-chemical structure of disk atmospheres in this way, and if higher accretion rates and higher rates of mechanical heating derive from the same physical mechanism, we would expect to see a correlation between accretion rate and H$_{2}$O feature strength. Accretion rate may play a similar role in enhancing HCN emission strength, i.e. by increasing the column density of warm HCN in the disk atmosphere. 

There may be an additional chemical connection between H$_{2}$O and HCN emission, with efficient water formation possibly leading to an enhanced HCN abundance.  As described by Lahuis \& van Dishoeck (2000), efficient H$_2$O formation will drive most of the available oxygen into H$_2$O, resulting in a lower abundance of gaseous O$_2$.  Since O$_2$ would otherwise react with atomic carbon, the lack of O$_2$ could lead to an enhanced atomic C abundance and in turn a larger HCN abundance (e.g., via the reaction scheme described by Agundez et al.\ 2008).
Perhaps for this reason, hot cores that are found to have the highest gas phase H$_{2}$O abundances are also those with the highest HCN abundances (e.g., van Dishoeck 1998; Lahuis \& van Dishoeck 2000). Thus accretion-related mechanical heating in disks may enhance disk HCN emission both thermally, by producing a deeper temperature inversion at the disk surface, and chemically, by enhancing the HCN abundance as a consequence of efficient water formation. Detailed modeling is needed to explore these possibilities.

Increased UV irradiation produced by higher stellar accretion may also enhance the HCN abundance. Using Ag{\'u}ndez et al.\ (2008) as a guide, 
Pascucci et al.\ (2009) argued that the HCN abundance in disk atmospheres 
may be limited by the availability of atomic nitrogen and that the atomic nitrogen 
abundance depends primarily on the dissociation of N$_2$ via UV-dissociation.
Thus, HCN would be brighter for sources with more energetic UV flux (i.e., higher accretion rate), while C$_2$H$_2$ (not requiring nitrogen to form) would not vary. This may explain their finding that T Tauri stars have stronger HCN emission relative to C$_2$H$_2$ than lower mass stars and brown dwarfs, as these lower mass objects would have lower photospheric UV emission and lower accretion rates than TTS. The range in stellar accretion rate among T Tauri stars may induce a range in their HCN abundances for similar reasons. 

Another factor that may play a role in setting the HCN flux from the disk is X-ray irradiation, based on Fig.\,\ref{fig_fp}b. The effect of X-ray irradiation on the thermal-chemical structure of disks has been investigated previously by Glassgold et al.\ (2004, 2009), although they did not specifically study HCN.\, X-ray irradiation may enhance the abundance of molecular ions and radicals that lead to enhanced HCN emission. Further modeling is needed to investigate the relative roles of X-ray and UV irradiation in this context.

We find a possible trend of HCN flux decreasing with stellar spectral type (Fig.\,\ref{fig_fp}c). While this is in the spirit of the trend found by Pascucci et al.\ (2009), it is unlikely that stellar spectral type itself (i.e., stellar temperature) is affecting the HCN flux for this small sample of TTS. The other two processes we examined, stellar accretion rate and stellar X-ray flux (and/or other processes not yet identified) are likely to have a more direct influence on the HCN flux. Stellar accretion rate is not well correlated with spectral type (see Fig.\,\ref{fig_fp}e) and the TTS in our sample span a small mass range, so the resulting accretion luminosity seems unlikely to be correlated over the range of spectral types that we studied. In comparison, L$_{\rm{X}}$ shows a possible correlation with spectral type (Fig.\,\ref{fig_fp}d), so it may be responsible for the moderate correlation of HCN flux with spectral type.

Several of the objects in our sample (plotted as open triangles in Fig.\,\ref{fig_fp}a, b, and c) appear to deviate from the possible trends we identify here. The dispersion
we observe could arise from differences in disk structure (e.g., flaring) and composition that may originate from the natal environment as well as the dynamic processing that occurs within the disk lifetime. This makes the objects that deviate from our observed trends not only expected, but of particular interest. For example, while variations in stellar accretion rate are typically factors of $\sim$\,2 or less (Hartigan et al.\ 1991),  stellar accretion rates of some individual sources may
vary up to an order of magnitude on timescales of $\sim$\,1 yr (Alencar \& Batalha 2002). This could induce a significant shift for some objects in our plots. 
Variability in the stellar accretion rate could also affect the time-averaged disk chemistry. Similar considerations might apply for stellar X-ray variability.

Another potential cause of dispersion is a different or additional 
heating source. The strength of the UV irradiation striking the disk may depend on the absorption along 
the line-of-sight, e.g., in a magnetosphere or an intervening wind 
(e.g., Alexander et al.\ 2004; Ercolano et al.\ 2008,
2009; Gorti \& Hollenbach 2008, 2009). This could influence the
temperature and chemical processing of the disk atmosphere, as might
radial transport or vertical mixing between the upper layer and
regions closer to the disk midplane (e.g., Bergin et al.\ 2007;
Turner et al.\ 2006; Semenov et al.\ 2006; Willacy et al.\ 2006). 

Dust sedimentation can also increase the line-to-continuum contrast of molecular emission (Glassgold et al.\ 2004; Dullemond et al.\ 2007), and such emission is more commonly detected in more highly settled disks (Salyk et al.\,2011). The properties and distribution of grains are known to vary widely
over disk age and structure (e.g., Watson et al.\ 2009). If molecular
formation (e.g., H$_{2}$) on grains influences disk chemical synthesis, variations
in grain properties may lead to variations in observable molecular
features (Glassgold et al.\ 2009). In the panels of Figure \ref{fig_fp},
there are several outlying points whose HCN flux index is enhanced or
depleted relative to the rest of the points. These might be
ideal systems in which to look for additional chemical peculiarity
or heating mechanisms that could be affecting the molecular emission
strength.

The trends described here require a larger sample to confirm. In tandem, it may be possible to expand the wavelength range we analyze by considering observations from \textit{Spitzer} IRS modules that cover a wider wavelength range (i.e., Long-High, 20$\,\mu$m$-$40$\,\mu$m) and more molecular species. Additional high resolution data would also help verify the technique of using SL spectra to recover real trends. 

\section{Summary \& Conclusions}

Our goal was to investigate the extent to which lower
resolution \textit{Spitzer} IRS data can be used to recover quantitative molecular emission trends 
seen in higher resolution \textit{Spitzer}
IRS data. 
We have shown that a simple prescription for measuring the strength 
of the 14 $\mu$m HCN emission feature, when applied to low resolution 
\textit{Spitzer} data, can recover trends in HCN emission strength that 
are seen in high resolution \textit{Spitzer} data. Additionally, we report possible correlations between HCN flux and stellar
accretion rate, and HCN flux and stellar X-ray luminosity, that may 
originate from accretion-driven 
mechanical heating and/or photochemistry at work in
the inner disk atmosphere. While qualitative comparisons of the presence of line emission were possible and successful earlier (e.g., Pascucci et al.\ 2009), our results demonstrate that quantitative comparisons of the line intensities can also be carried out.
 
What controls the presence and strength of organic molecular features
such as HCN in the planet-forming regions around young stars?
One challenge in addressing this question 
is the large number of physical and chemical processes 
that can potentially affect the molecular emission strength, as discussed in \S 4. 
Our methods and results show that the large number of low resolution disk spectra that reside in the \textit{Spitzer} archive could be used in future demographic 
studies to attempt to identify the relevant processes. 

\acknowledgments

{\it Facilities:} \facility{Spitzer ()}


\clearpage
\begin{thebibliography}{dummy}

\bibitem[Ag{\'u}ndez et al.(2008)]{2008A&A...483..831A} Ag{\'u}ndez, M., Cernicharo, J., \& Goicoechea, J.~R.\ 2008, \aap, 483, 831 

\bibitem[Alencar \& Batalha(2002)]{2002ApJ...571..378A} Alencar, S.~H.~P., \& Batalha, C.\ 2002, \apj, 571, 378 

\bibitem[Alexander et al.(2004)]{2004MNRAS.354...71A} Alexander, R.~D., Clarke, C.~J., \& Pringle, J.~E.\ 2004, \mnras, 354, 71 

\bibitem[Apai \& Lauretta(2010)]{2010pdac.book....1A} Apai, D., \& Lauretta, D.~S.\ 2010, Protoplanetary Dust: Astrophysical and Cosmochemical Perspectives, 128 

\bibitem[Ardila \& Basri(2000)]{2000ApJ...539..834A} Ardila, D.~R., \& Basri, G.\ 2000, \apj, 539, 834 

\bibitem[Bergin et al.(2007)]{2007prpl.conf..751B} Bergin, E.~A., Aikawa, Y., Blake, G.~A., \& van Dishoeck, E.~F.\ 2007, Protostars and Planets V, 751 

\bibitem[Bergin et al.(2004)]{2004ApJ...614L.133B} Bergin, E., et al.\ 2004, \apjl, 614, L133 

\bibitem[Berthoud et al.(2007)]{2007ApJ...660..461B} Berthoud, M.~G., Keller, L.~D., Herter, T.~L., Richter, M.~J., \& Whelan, D.~G.\ 2007, \apj, 660, 461 

\bibitem[Blake \& Boogert(2004)]{2004ApJ...606L..73B} Blake, G.~A., \& Boogert, A.~C.~A.\ 2004, \apjl, 606, L73 

\bibitem[Blum et al.(2004)]{2004ApJ...617.1167B} Blum, R.~D., Barbosa, C.~L., Damineli, A., Conti, P.~S., \& Ridgway, S.\ 2004, \apj, 617, 1167  

\bibitem[Brittain et al.(2007)]{b07} Brittain, S.~D., Simon, T., Najita, J.~R., \& Rettig, T.~W.\ 2007, \apj, 659, 685 

\bibitem[Carmona(2010)]{2010EM&P..tmp....1C} Carmona, A.\ 2010, Earth Moon and Planets, 106, 71 

\bibitem[Carr \& Najita (2011)]{cn11} Carr, J.S., \& Najita, J.R. 2011, ApJ, submitted

\bibitem[Carr \& Najita (2008)]{cn08} Carr, J.S., \& Najita, J.R. 2008, Science, 319, 1504

\bibitem[Carr(2005)]{2005hris.conf..203C} Carr, J.\ 2005, High Resolution Infrared Spectroscopy in Astronomy, 203 

\bibitem[Carr et al.(2004)]{2004ApJ...603..213C} Carr, J.~S., Tokunaga, A.~T., \& Najita, J.\ 2004, \apj, 603, 213 

\bibitem[Carr et al.(2001)]{2001ApJ...551..454C} Carr, J.~S., Mathieu, R.~D., \& Najita, J.~R.\ 2001, \apj, 551, 454 

\bibitem[Carr et al.(1993)]{1993ApJ...411L..37C} Carr, J.~S., Tokunaga, A.~T., Najita, J., Shu, F.~H., \& Glassgold, A.~E.\ 1993, \apjl, 411, L37 

\bibitem[Chandler et al.(1993)]{1993ApJ...412L..71C} Chandler, C.~J., Carlstrom, J.~E., Scoville, N.~Z., Dent, W.~R.~F., \& Geballe, T.~R.\ 1993, \apjl, 412, L71 

\bibitem[Dullemond et al.(2007)]{2007prpl.conf..555D} Dullemond, C.~P., Hollenbach, D., Kamp, I., \& D'Alessio, P.\ 2007, Protostars and Planets V, 555 

\bibitem[Dutrey et al.(2007)]{dgh07} Dutrey, A., Guilloteau, S., \& Ho, P.\ 2007, Protostars and Planets V, 495 


\bibitem[Dutrey et al.(1998)]{1998A&A...338L..63D} Dutrey, A., Guilloteau, S., Prato, L., Simon, M., Duvert, G., Schuster, K., \& Menard, F.\ 1998, \aap, 338, L63 


\bibitem[Dutrey et al.(1996)]{1996A&A...309..493D} Dutrey, A., Guilloteau, S., Duvert, G., Prato, L., Simon, M., Schuster, K., \& Menard, F.\ 1996, \aap, 309, 493 


\bibitem[Ercolano et al.(2009)]{2009ApJ...699.1639E} Ercolano, B., Clarke, C.~J., \& Drake, J.~J.\ 2009, \apj, 699, 1639 

\bibitem[Ercolano et al.(2008)]{2008ApJ...688..398E} Ercolano, B., Drake, J.~J., Raymond, J.~C., \& Clarke, C.~C.\ 2008, \apj, 688, 398 

\bibitem[Furlan et al.(2006)]{f06} Furlan, E., et al.\ 2006, \apjs, 165, 568 

\bibitem[Glassgold et al.(2009)]{gmn09} Glassgold, A.~E., Meijerink, R., \& Najita, J.~R.\ 2009, \apj, 701, 142 

\bibitem[Glassgold et al.(2004)]{gni04} Glassgold, A.~E., Najita, J., \& Igea, J.\ 2004, \apj, 615, 972 

\bibitem[Gorti \& Hollenbach(2009)]{2009ApJ...690.1539G} Gorti, U., \& Hollenbach, D.\ 2009, \apj, 690, 1539 

\bibitem[Gorti \& Hollenbach(2008)]{gh08} Gorti, U., \& Hollenbach, D.\ 2008, \apj, 683, 287 

\bibitem[G{\"u}del et 
al.(2010)]{2010A&A...519A.113G} G{\"u}del, M., et al.\ 2010, \aap, 519, A113 

\bibitem[Guilloteau \& Dutrey(1998)]{1998A&A...339..467G} Guilloteau, S., \& Dutrey, A.\ 1998, \aap, 339, 467 

\bibitem[Gullbring et al.(1998)]{g98} Gullbring, E., Hartmann, L., Briceno, C., \& Calvet, N.\ 1998, \apj, 492, 323 

\bibitem[Hartigan et al.(1994)]{h94} Hartigan, P., Strom, K.~M., \& Strom, S.~E.\ 1994, \apj, 427, 961 

\bibitem[Hartigan et al.(1991)]{1991ApJ...382..617H} Hartigan, P., Kenyon, S.~J., Hartmann, L., Strom, S.~E., Edwards, S., Welty, A.~D., \& Stauffer, J.\ 1991, \apj, 382, 617 

\bibitem[Hartmann et al.(1998)]{h98} Hartmann, L., Calvet, N., Gullbring, E., \& D'Alessio, P.\ 1998, \apj, 495, 385 

\bibitem[Henning \& Meeus(2009)]{2009arXiv0911.1010H} Henning, T., \& Meeus, G.\ 2009, arXiv:0911.1010 

\bibitem[Herczeg et al.(2006)]{2006ApJS..165..256H} Herczeg, G.~J., Linsky, J.~L., Walter, F.~M., Gahm, G.~F., \& Johns-Krull, C.~M.\ 2006, \apjs, 165, 256 

\bibitem[Herczeg et al.(2002)]{2002ApJ...572..310H} Herczeg, G.~J., Linsky, J.~L., Valenti, J.~A., Johns-Krull, C.~M., \& Wood, B.~E.\ 2002, \apj, 572, 310 

\bibitem[Houck et al.(2004)]{2004ApJS..154...18H} Houck, J.~R., et al.\ 
2004, \apjs, 154, 18 

\bibitem[Kastner et al.(1997)]{1997Sci...277...67K} Kastner, J.~H., Zuckerman, B., Weintraub, D.~A., \& Forveille, T.\ 1997, Science, 277, 67 

\bibitem[Kenyon \& Hartmann (1995)]{kh95} Kenyon, S.J., \& Hartmann, L. 1995, \apjs, 101, 117

\bibitem[Lahuis et al.(2006)]{l06} Lahuis, F., et al.\ 2006, \apjl, 636, L145 

\bibitem[Lahuis \& van Dishoeck(2000)]{2000A&A...355..699L} Lahuis, F., \& van Dishoeck, E.~F.\ 2000, \aap, 355, 699 

\bibitem[Millan-Gabet et al.(2007)]{2007prpl.conf..539M} Millan-Gabet, R., Malbet, F., Akeson, R., Leinert, C., Monnier, J., \& Waters, R.\ 2007, Protostars and Planets V, 539 

\bibitem[Muzerolle et al.(2005)]{2005ApJ...625..906M} Muzerolle, J., Luhman, K.~L., Brice{\~n}o, C., Hartmann, L., \& Calvet, N.\ 2005, \apj, 625, 906 

\bibitem[Najita et al.(2009)]{2009ApJ...691..738N} Najita, J.~R., Doppmann, G.~W., Carr, J.~S., Graham, J.~R., \& Eisner, J.~A.\ 2009, \apj, 691, 738 

\bibitem[Najita et al.(2008)]{2008ApJ...687.1168N} Najita, J.~R., Crockett, N., \& Carr, J.~S.\ 2008, \apj, 687, 1168 

\bibitem[Najita et al.(2007a)]{ncgv07} Najita, J.~R., Carr, J.~S., Glassgold, A.~E., \& Valenti, J.~A.\ 2007a, Protostars and Planets V, 507 

\bibitem[Najita et al.(2007b)]{nsm07} Najita, J.~R., Strom, S.~E., \& Muzerolle, J.\ 2007b, \mnras, 378, 369 

\bibitem[Najita et al.(2003)]{ncm03} Najita, J., Carr, J.~S., \& Mathieu, R.~D.\ 2003, \apj, 589, 931 

\bibitem[Najita et al.(2000)]{2000prpl.conf..457N} Najita, J.~R., Edwards, S., Basri, G., \& Carr, J.\ 2000, Protostars and Planets IV, 457 

\bibitem[Najita et al.(1996)]{1996ApJ...462..919N} Najita, J., Carr, J.~S., Glassgold, A.~E., Shu, F.~H., \& Tokunaga, A.~T.\ 1996, \apj, 462, 919 

\bibitem[Natta et al.(2007)]{2007prpl.conf..767N} Natta, A., Testi, L., Calvet, N., Henning, T., Waters, R., \& Wilner, D.\ 2007, Protostars and Planets V, 767 

\bibitem[Pascucci et al.(2009)]{p09} Pascucci, I., Apai, D., Luhman, K., Henning, T., Bouwman, J., Meyer, M.~R., Lahuis, F., \& Natta, A.\ 2009, \apj, 696, 143 

\bibitem[Preibisch et al.(2005)]{2005ApJS..160..401P} Preibisch, T., et 
al.\ 2005, \apjs, 160, 401 

\bibitem[Pontoppidan et al.(2010)]{2010arXiv1006.4189P} Pontoppidan, K.~M., 
Salyk, C., Blake, G.~A., Meijerink, R., Carr, J.~S., 
\& Najita, J.\ 2010, arXiv:1006.4189 

\bibitem[Pontoppidan et al.(2008)]{2008ApJ...684.1323P} Pontoppidan, K.~M., 
Blake, G.~A., van Dishoeck, E.~F., Smette, A., Ireland, M.~J., 
\& Brown, J.\ 2008, \apj, 684, 1323 

\bibitem[Qi et al.(2008)]{2008ApJ...681.1396Q} Qi, C., Wilner, D.~J., Aikawa, Y., Blake, G.~A., \& Hogerheijde, M.~R.\ 2008, \apj, 681, 1396 

\bibitem[Salyk et al.(2009)]{s09} Salyk, C., Pontoppidan, K.~M., Blake, G.~A., Najita, J., \& Carr, J. 2011, \apj, accepted

\bibitem[Salyk et al.(2009)]{s09} Salyk, C., Blake, G.~A., Boogert, A.~C.~A., \& Brown, J.~M.\ 2009, \apj, 699, 330 

\bibitem[Salyk et al.(2008)]{s08} Salyk, C., Pontoppidan, K.~M., Blake, G.~A., Lahuis, F., van Dishoeck, E.~F.,  \& Evans, N.~J., II 2008, \apjl, 676, L49 

\bibitem[Salyk et al.(2007)]{2007ApJ...655L.105S} Salyk, C., Blake, G.~A., Boogert, A.~C.~A., \& Brown, J.~M.\ 2007, \apjl, 655, L105 

\bibitem[Semenov et al.(2006)]{2006ApJ...647L..57S} Semenov, D., Wiebe, D., \& Henning, T.\ 2006, \apjl, 647, L57 

\bibitem[Semenov et al.(2005)]{2005ApJ...621..853S} Semenov, D., Pavlyuchenkov, Y., Schreyer, K., Henning, T., Dullemond, C., \& Bacmann, A.\ 2005, \apj, 621, 853 

\bibitem[Stelzer \& Neuh{\"a}user(2001)]{2001A&A...377..538S} Stelzer, B., \& Neuh{\"a}user, R.\ 2001, \aap, 377, 538 

\bibitem[Stone et al.(2000)]{s00} Stone, J.~M., Gammie, C.~F., Balbus, S.~A., \& Hawley, J.~F.\ 2000, Protostars and Planets IV, 589 

\bibitem[Telleschi et al.(2007)]{2007A&A...468..425T} Telleschi, A., G{\"u}del, M., Briggs, K.~R., Audard, M., \& Palla, F.\ 2007, \aap, 468, 425 

\bibitem[Thi \& Bik(2005)]{2005A&A...438..557T} Thi, W.-F., \& Bik, A.\ 2005, \aap, 438, 557

\bibitem[Thi et al.(2005)]{2005A&A...430L..61T} Thi, W.-F., van Dalen, B., Bik, A., \& Waters, L.~B.~F.~M.\ 2005, \aap, 430, L61 

\bibitem[Thi et al.(2004)]{2004A&A...425..955T} Thi, W.-F., van Zadelhoff, G.-J., \& van Dishoeck, E.~F.\ 2004, \aap, 425, 955 

\bibitem[Turner et al.(2006)]{2006ApJ...639.1218T} Turner, N.~J., Willacy, K., Bryden, G., \& Yorke, H.~W.\ 2006, \apj, 639, 1218 

\bibitem[Valenti et al.(2000)]{2000ApJS..129..399V} Valenti, J.~A., Johns-Krull, C.~M., \& Linsky, J.~L.\ 2000, \apjs, 129, 399 

\bibitem[van Dishoeck(1998)]{1998ISAA....4...53V} van Dishoeck, E.~F.\ 
1998, The Molecular Astrophysics of Stars and Galaxies, edited by Thomas 
W.~Hartquist and David A.~Williams.~Clarendon Press, Oxford, 1998., p.53, 
4, 53 

\bibitem[Watson et al.(2009)]{w09} Watson, D.~M., et al.\ 2009, \apjs, 180, 84 

\bibitem[Willacy et al.(2006)]{2006ApJ...644.1202W} Willacy, K., Langer, W., Allen, M., \& Bryden, G.\ 2006, \apj, 644, 1202 

\bibitem[Winston et al.(2010)]{2010AJ....140..266W} Winston, E., et al.\ 2010, \aj, 140, 266 

\end{thebibliography}
\end{document}